\documentclass[12pt]{article}
\usepackage{graphicx}
\usepackage{subfigure}
\usepackage{cite}
\usepackage{amsmath}
\usepackage{capt-of}

\usepackage{tikz}
\usetikzlibrary{positioning,arrows.meta,shapes}

\begin{document}

\title{\vspace{-2cm} Two stroke Pumping Technique  \\ for \\ Many-Body Systems}

\author{ Serge Galam\thanks{serge.galam@sciencespo.fr} \\
CEVIPOF - Centre for Political Research, Sciences Po and CNRS,\\
1, Place Saint Thomas d'Aquin, Paris 75007, France}

\date{Entropy 28(2), 202 (2026)\\
https://doi.org/10.3390/e28020202}

\maketitle

\begin{abstract}

I introduce a new analytical framework for estimating critical temperatures in interacting many-body systems, focusing on the Ising model. Combining the Bethe cluster setting, the Metropolis update, and the Galam Majority Model developed in sociophysics, I build a two stroke pumping technique (TSP). Applied to the Ising model in dimensions $d=2, 3, 4$, TSP yields values of $T_c$ which are all at an excess of $+0.03$ from exact estimates. At $d=1$ the exact value $T_c=0$ is obtained. In addition, TSP indicates analytically the practical impossibility to reach full symmetry breaking at $T=0$. The results are thus found in good agreement with numerical findings while requiring significantly fewer computational resources than Monte Carlo sampling. Calculations are computationally efficient and transparent.  The framework is general and can be extended to a broad class of discrete spin models. This positions TSP as an intermediate yet scalable tool for studying cooperative behavior in many body interacting systems.

\end{abstract}


\section{Introduction}

Understanding the collective behavior of systems composed of many interacting degrees of freedom remains a central focus in statistical physics and beyond. In particular, since even simple local rules may exhibit rich and sometimes unexpected emergent collective phenomena. With this regard, the Ising model, introduced nearly a century ago \cite{Ising1925}, has played and is still playing, a central driving role to tackle the issue.

Despite its minimal definition, consisting of binary spins $s_i = \pm 1$ on a lattice with nearest-neighbor coupling, it captures most of the essential features of cooperative ordering, critical behavior, and universality. The model's conceptual simplicity and rich phenomenology have made it the standard testing ground for both analytical approximations and numerical simulations aimed at elucidating how macroscopic properties emerge from short and long range microscopic interactions.

Over the decades, thanks to its simplicity and generality, the Ising model became a cornerstone prototype to investigate many body problems. The model provides a unifying framework for studying how collective behavior arises from local interactions in many systems with discrete symmetry, far beyond its original magnetic background. 

Indeed, the same mathematical structure has been used to describe cooperative phenomena in a wide variety of systems, including binary alloys and lattice gases \cite{Brush1967}, neural networks \cite{MezardMontanari2009}, protein folding and glassy dynamics \cite{Kadanoff2000}, as well as structural phase transitions and adsorption phenomena on surfaces. 

Moreover, numerous Ising-like formulations have been also successfully adapted  out the traditional realm of physics, to model decision-making processes, opinion dynamics and social phenomena, where interacting agents replace interacting spins \cite{n1, n2, p6, p7, p8}. 

These interdisciplinary extensions have clarified how macroscopic collective behavior can emerge from local microscopic interactions even in complex or heterogeneous social networks, similarly to condensed matter physics. The large spectrum of applications of the Ising model has reinforced its framework as one of the most versatile and influential paradigms in modern quantitative sciences.

This diversity of applications underscores the model's fundamental nature and motivates the pursuit of accurate theoretical and numerical methods capable of capturing its critical behavior. Understanding how the critical temperature and other macroscopic properties depend on the dimensionality and connectivity of the underlying lattice remains a central challenge in both physics and related disciplines. In particular, evaluation of the critical exponents is instrumental to describe a related phase transition.

Thus, determining the critical temperature $T_c$ as a function of the various parameters of the model is an instrumental challenge.  Especially since up to date the Ising model has been solved exactly only in one and two dimensions.

In one dimension, the model does not exhibit a finite-temperature phase transition with no order occurring at $T>0$. However, in two dimensions, the Onsager exact solution yields a finite $T_c$ \cite{Onsager1944}. In three and higher dimensions, the model being not exactly solvable, approximate or numerical approaches must be used. 

With respect to analytical approximations, mean-field theory provides the simplest treatment, but it overestimates $T_c$ because it neglects fluctuations entirely. However, deviations from exact and numerical estimates decrease with increasing dimension, as fluctuations are gradually averaged out.

The Bethe approximation improves on the mean field result by incorporating short-range correlations via a tree-like effective lattice, yielding more accurate estimates of $T_c$, particularly in low-dimensional systems \cite{Bethe1935}. Nevertheless, it still fails to account for loop contributions that are crucial in finite-dimensional lattices.

Later, the development of the renormalization group (RG) theory has marked a major analytical breakthrough in the understanding of phase transitions and critical phenomena \cite{Wilson1971, Fisher1974, Ma1976, Kadanoff2000}. However, implementing RG transformations requires careful coarse-graining, truncation of degrees of freedom, and often intricate perturbative expansions, making the analysis technically more demanding than direct numerical simulations \cite{Wilson1974,Fisher1998}.

At the current stage of available analytical tools, to get accurate estimates of critical temperatures at any dimension requires running Monte Carlo simulations using either Metropolis or Glauber dynamics  \cite{Metropolis1953, Glauber1963}. But then,  large size systems are needed making high the computational cost  \cite{mc1}.

Last, but not least, an had hoc formula for Ising critical temperatures inspired from another had hoc formula for percolation thresholds, was shown to yield very good estimates of Ising $T_c$ at all dimensions \cite{GalamMauger1996}. But till now, no derivation has been found.

Therefore, there is a methodological gap between analytically tractable but oversimplified approximations (mean-field and Bethe) and computationally heavy but accurate simulations (Monte Carlo, numerical RG). The goal of this work is thus to introduce a new simple analytical scheme to reduce this gap. 

Accordingly, I build a novel scheme, which combines the Bethe cluster setting, the Metropolis update, and the Galam Majority Model (GMM) \cite{GMM1, GMM2} developed in sociophysics  \cite{r1, r2, r3}. A two stroke pumping technique is thus obtained (TSP). When applied to the Ising model in $d$-dimensions, associated values of $T_c$ can be extracted. The results are found to be in good agreement with exact and numerical estimates while requiring very little computational resources. In addition, while TSP recovers the exact $T_c=0$ at $d=1$, it also indicates the practical impossibility to reach full symmetry breaking at $T=0$.

Indeed, a previous work has applied the Global Unifying Frame (GUF)  \cite{guf} to the 2-dimensional Ising system using both Metropolis and Glauber updates \cite{GalamMartins}. While the associated dynamics is different from the two stroke pumping technics, the associated value of $T_c$ is identical.

The rest of this article is organized as follows. I review the Ising model and various scheme to calculate its critical temperature at various dimensions using mean field, Bethe, real-space renormalization group, series, Monte Carlo and had hoc Galam-Mauger (GM) formula. The two stoke pumping technique is presented in Section 3 and implemented at $d=2$. Section 4 deals with applying TSP at $d=1$, Section 5 to $d=3$ and Section 6 to $d=4$. Concluding remarks are presented in the Conclusion.

\section{The Ising model}

Before introducing my new proposal to evaluate the critical temperatures of the Ising model defined on a $d$-dimensional hypercubic lattice,  it is useful to recall its basic setting. 

Given a lattice, each site carries a spin variable $s_i=\pm 1$. All the spins interact via nearest-neighbor interactions with a coupling constant $J$ where $J>0$ favors ferromagnetic ordering with parallel alignment, and $J<0$ antiferromagnetic ordering with anti-parallel alignment. The Hamiltonian reads,
\begin{equation}
\mathcal{H} = - J \sum_{\langle i j \rangle} s_i s_j -h  \sum_{i} s_i - \sum_{i} h_i s_i   ,
\label{H}
\end{equation}
where $\langle ij\rangle$ runs over all distinct nearest-neighbor pairs of the lattice, $h$ is an external uniform field applied to all spins, and $h_i$ is a local field, which couples linearly to the spin $s_i$ at site $i$ and may vary from site to site. It can positive, negative or zero.

While the coupling favors ordering, a non-zero temperature $T\neq 0$ produces thermal fluctuations, which in turn creates disorder. Thermal fluctuations are governed by the Boltzmann weight $\exp(-\mathcal{H}/T)$, where the temperature $T$ is expressed in units with $k_B=1$.

For the hypercube in dimensions $d$, the coordination number is $z=2d$.  In the absence of both uniform ($h=0$) and local
fields ($h_i=0$), the model exhibits a thermal phase transition for all dimensions $d\ge 2$.  

At high temperature, thermal agitation dominates and the system remains disordered with zero magnetization.  However, at  low temperatures, the interaction term prevails on the thermal fluctuations and the system spontaneously orders, developing a nonzero magnetization.

The temperature separating these two regimes is the critical temperature $T_c(d)$, or equivalently the critical coupling $K_c(d)=1/T_c(d)$ when setting $J=k_B=1$.  Determining $T_c(d)$ is a central problem in statistical physics. It is known exactly only in $d=1$ with $T_c=0$  (no finite-temperature transition) and $d=2$ with Onsager's solution. For $d\ge 3$ the critical temperatures must be estimated through approximations including mean-filed and Bethe, high-temperature series, renormalization-group methods, or large-scale Monte Carlo simulations. The dependence of $T_c$ on dimension provides a direct measure of how geometry and connectivity shape the balance between order and disorder in interacting spin systems,

Before moving on to my new proposal to evaluate critical temperatures of the nearest-neighbor Ising model at dimensions $d$, I report current available estimates obtained using several different techniques. More precisely, I list a series of estimates for $K_c=1/T_c$ with coupling constant $J=1$ using respectively, mean-field theory, Bethe approximation, real-space renormalization using the Migdal-Kadanoff, high-temperature series expansions, monte Carlo simulations, combining finite-size scaling and high-statistics sampling, and the GM empirical formula.

\subsection{Mean-field (MF)}

Mean-field (MF) estimates assume that each spin experiences an effective field proportional to the average magnetization. This yields
\begin{equation}
K_c = \frac{1}{z} ,
\label{mf}
\end{equation}
where $z$ is the coordination number \cite{Stanley1971}. MF theory predicts wrongly a transition at $d=1$ but correctly a continuous transition at $d\leq 2$. However, as MF neglects spatial correlations, related $T_c$ are systematically overestimated at low dimensions, i.e., underestimating $K_c$  as seen in Table \ref{Kcs}. For example, on a square lattice ($z=4$), MF predicts $K_c=O.250000$, nearly half the exact Onsager value. In three dimensions ($z=6$), $K_c=0.166667$, compared to the Monte Carlo value of about $0.221655$ \cite{Huang1987}. MF becomes exact when the lattice is fully connected.

\subsection{Bethe }

The Bethe approximation \cite{Bethe1935} treats a central spin exactly while assuming that its z neighbors are uncorrelated and each has a magnetization equals to the mean magnetization. Within this approximation, the critical coupling is,
\begin{equation}
 K_c^{\rm Bethe} = \operatorname{atanh}\frac{1}{z-1} .
\label{bethe}
\end{equation}

For a one-dimensional chain ($z=2$), Bethe gives $K_c^{\rm Bethe} \to \infty$, corresponding to $T_c = 0$. For a square lattice ($z=4$), it yields $K_c \approx 0.346574$, underestimating the exact Onsager value $K_c \approx 0.440687$ \cite{Onsager1944}. With respect to MF, Bethe theory incorporates local correlations and partially accounts for dimensional effects \cite{Baxter1982}. Results are better from $d=4$ and up.

\subsection{Real-space renormalization (RG)}

Real-space renormalization in the Migdal--Kadanoff (MK) approach \cite{Migdal1975, Kadanoff1976} relies on a coarse-graining transformation combining bond moving and decimation to approximate the exact real-space RG flow of the Ising model. Despite its simplicity, this coarse-graining scheme yields a coherent sequence of critical couplings for the hypercubic Ising model across dimensions as seen in Table \ref{Kcs}. These values reflect the typical accuracy of MK coarse-graining, which are qualitatively correct in low dimensions and increasingly accurate at higher dimensions \cite{NauenbergScalapino1974, BerkerOstlund1979}.

\subsection{Series}

High-temperature series expansions \cite{Domb1960,Butera2002} give precise estimates of $K_c$ by summing contributions from clusters of increasing size and using Pad\'e extrapolations. They converge to the exact $d=2$ solution \cite{Onsager1944} and to high-precision Monte Carlo results in $d \ge 3$\cite{Blote1995,Hasenbusch2010,DengBlote2003,Landau2018,LundowMarkstrom2009,LundowMarkstrom2011}. Series methods explicitly capture both short- and long-range correlations, accurately reflecting the dimensional dependence of $T_c$.

\subsection{Monte Carlo simulations (MC)}

Monte Carlo simulations, combining finite-size scaling and high-statistics sampling, yield some of the most precise at $d=2, 3, 4,5$ \cite{Blote1995,LundowMarkstrom2009,LundowMarkstrom2011}. 

Monte Carlo simulations, using Metropolis \cite{Metropolis1953} or Glauber dynamics \cite{Glauber1963}, provide numerical benchmarks that reproduce known critical temperatures with high precision. For $d=2$, Monte Carlo reproduces Onsager's exact value. These methods directly sample equilibrium distributions, capturing full many-body correlations, but are affected by critical slowing down near $T_c$, necessitating finite-size scaling \cite{Binder1981}, histogram reweighing \cite{Ferrenberg1988} and very large sample sizes to achieve high accuracy.

\subsection{Galam--Mauger  formula (GM)}

Finally, the ad hoc GM formula \cite{GalamMauger1996} writes,
\begin{equation}
 K_c = K_0 \Bigl[(1-\frac{1}{d})(2d-1)\Bigr]^{-a}, 
 \label{GM}
\end{equation}
with  $K_0=0.626036$ and $a=0.863375$. The formula produces highly accurate estimates in all dimensions as seen in Table \ref{Kcs}. However, no analytical derivation has been found so far.

 \subsection{Table of $K_c$ estimates}

Table \ref{Kcs} shows tha associated estimates at dimensions $d=1, 2, 3, 4$. All values are given with 3 digits. At $d=1$  $K_C=\infty \Leftrightarrow T_c=0$ and at $d=2$ Onsager solution \cite{Onsager1944} writes,
\begin{equation}
K_c = \frac{1}{2}\ln(1+\sqrt{2}) \approx 0.440687.
\label{on}
\end{equation}

\begin{table}[h!]
\centering
\begin{tabular}{|c|c|c|c|c|c|c|c|c|}
\hline
$d$ & $z$ & MF & Bethe & RG & Series & MC & GM \\
\hline
1 & 2  & 0.500 & \(\infty\) & \(\infty\) & \(\infty\) & \(\infty\) & \(\infty\)  \\
2 & 4  & 0.250 & 0.347& 0.401 & 0.441 & 0.441 & 0.441 \\
3 & 6  & 0.167 & 0.203 & 0.215 & 0.222 & 0.222& 0.221 \\
4 & 8  & 0.125 & 0.144 & 0.148 & 0.150 & 0.150 & 0.150 \\
\hline
\end{tabular}
\caption{Critical coupling $K_c = 1/T_c$ of the nearest-neighbor Ising model ($J=1$, $k_B=1$), rounded to 3 significant digits for dimensions $d=1, 2, 3, 4$ and a number of respective numbers of nearest-neighbors $z=2, 4, 6, 8$ using mean-field (MF); Bethe; Real-space renormalization within Migdal-Kadanoff approach (RG); Series; Monte Carlo simulations (MC); GM formula.}
\label{Kcs}
\end{table}

\section{Two stroke pumping technique (TSP)}

A decade ago, with Martins we applied the Global Unifying Frame (GUF) developed in sociophysics to evaluate the critical temperature of the 2-d Ising model with groups of 5 spins to account for a central spin and its 4 nearest-neighbors  \cite{GalamMartins}. 

Given a number of interacting agents, here five, GUF enumerates all possible configurations of  \cite{guf}. An update of each configuration is applied according to the specificity of the model, here only the central spin was updated according to either a Metropolis or a Glauber rule leaving the other four unchanged. Then all agents are randomly reshuffled to obtain a new distribution within the same configurations. The process is repeated until an equilibrium state is obtained. A general sequential probabilistic frame is thus built from which a phase diagram can be constructed.

The Metropolis update yields a first order transition with two critical temperatures $T_{c1} \approx 1.59$ and $T_{c2}  \approx 2.11$, the second one being close to the exact Onsager result $T_c  \approx 2.27$. It is worth to notice that the vertical first-order transition line is reminiscent of the abrupt vanishing of the Onsager second-order transition. Surprisingly, the Glauber update yields a continuous transition at $T_{c}  \approx 3.09$ closer to the Bethe estimate of $2.88$ than to the MF value $T_c=4$.

On this basis, I extend the Galam-Martins scheme by building a two stroke pumping technique (TSP). Combining a Bethe topology, a single spin Monte Carlo update and the GMM iteration scheme, I obtain an analytical formula valid  at any dimension $d$. Here, I apply the formula at $d=1, 2, 3, 4$ to evaluate the associated critical temperatures.

\subsection{The 2-dimensional case}

I start at $d=2$ with a cluster of one spin and its 4 nearest-neighbors. For all five spins, the probability to have the spin up is $p_0$ and $(1-p_0)$ to have it down. Then, I update the central spin according to the Metropolis scheme applied to the actual configuration of the 4 nearest-neighbors spins, which are not updated. Accordingly, given a surrounding configuration for a given site $i$, the energy change associated with flipping the spin $s_i$ is,
\begin{equation}
 \Delta E
= 2 s_i E_c ,
 \label{M1}
\end{equation}
with
\begin{equation}
E_c\equiv -J \sum_{j=1}^4 s_j 
 \label{M2}
\end{equation}
where $j=1, 2, 3, 4$ denote the 4 nearest-neighbors of spin $s_i$ and $c$ labels one of its 16 surrounding possible configurations.  The proposed flip $s_i \to -s_i$ is then accepted with the Metropolis probability,
\begin{equation}
p_{\mathrm{M}} =
\begin{cases}
1, & \Delta E \le 0, \\[6pt]
\exp(-\Delta E/T), & \Delta E > 0 ,
\end{cases}
\label{M2}
\end{equation}
and rejected otherwise.

Following GMM, $\Delta E > 0 \Leftrightarrow s_i$ aligned with the local surrounding majority and a flip is done with probability $p_M$. When $\Delta E < 0 \Leftrightarrow s_i$ opposed to the majority and $\Delta E = 0 \Leftrightarrow$ there is a tie, a flip is systematically done.

Among the 16 configurations of the 4 nearest-neighbors spins, one ($c=1$) has 4 spins $+1$ and 0 spin $-1$, 4 ($c=2, 3, 4, 5$) have 3 spins $+1$ and one spin $-1$,  6 ($c=6, 7, 8, 9, 10, 11 $) have 2 spins $+1$ and 2 spins $-1$ (a tie),  4 ($c=12, 13, 14, 15$) have 1 spin $+1$ and 3 spins $-1$,  one ($c=16$) has 0 spin $+1$ and 4 spins $-1$.

The respective $E_c$ are $E_1= -4 J$, $E_{c=2,3,4,5}= -2J$,  $E_{c=6, 7, 8, 9, 10, 11}= 0$, $E_{c=12, 13, 14, 15}=2J$, $E_{c=16}=4J$. In addition, the associated probabilities of each group of configurations are $p^4$, $p^3(1-p)$, $p^2(1-p)^2$, $p(1-p)^3$, $(1-p)^4$.
 
\begin{figure}
\centering
\vspace*{-6.5cm}
\hspace*{-1.5cm}
\includegraphics[width=1.25\textwidth]{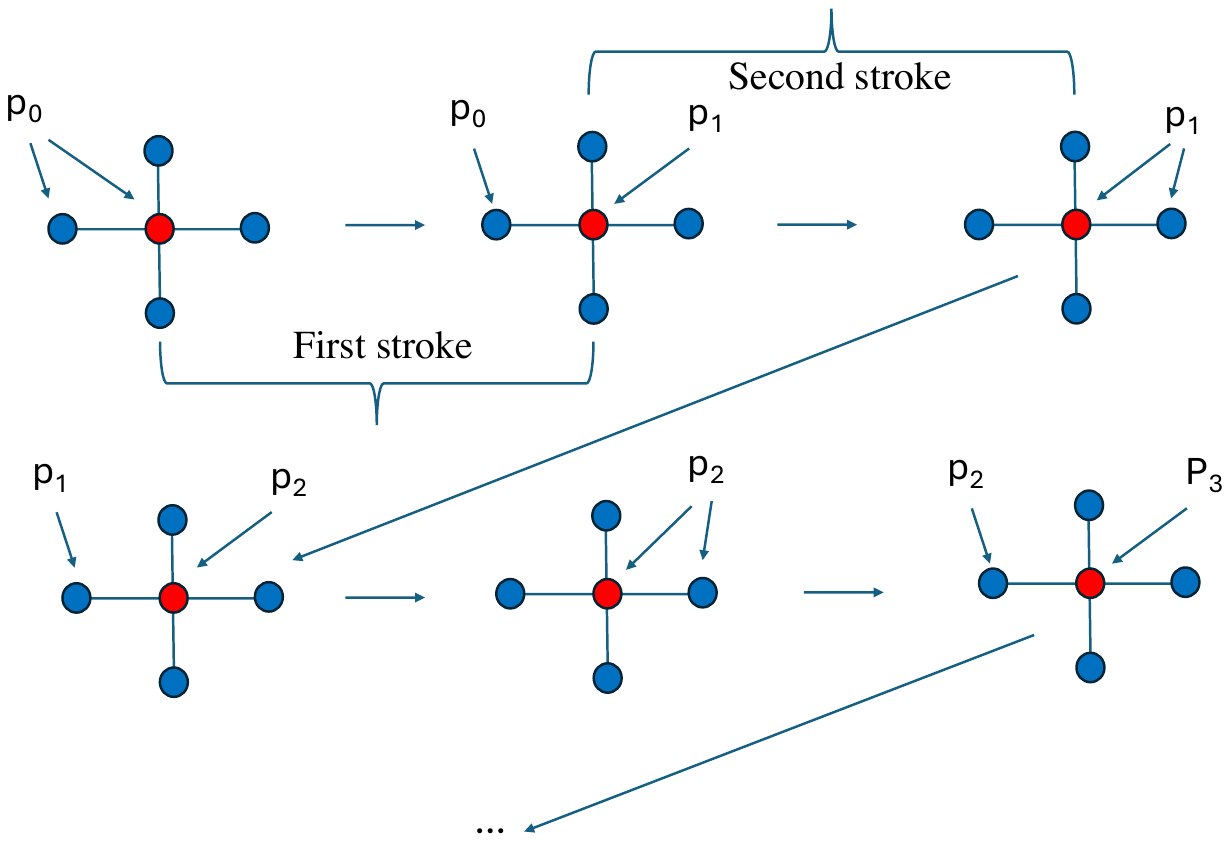}
\caption{Up left: a cluster of one spin and its 4 nearest-neighbors with probability $p_0$ to have $+1$ for each of them.  Up center: The central spin has been updated with a probability $p_1$ to be $+1$. That defines the first stroke of the pumping technique. Up right: the probability to $+1$ is now $p_1$ for all 5 spins defining the second stroke. Down left: the central spin is updated with probability $p_2$ to be $+1$. Down center: The probability to be $+1$ is $p_2$ for all 5 spins. Down right: the central spin is updated with probability $p_3$ to $+1$. and so forth till an attractor is reached.}
\label{tt1}
\end{figure}

Applying Eq. (\ref{M2}) to the spin $s_0$, which is equal to $+1$ with probability $p_0$, leads to a  new probability $p_1$ to have it equal to $+1$. This step is the first stroke of the new pumping technique. Then I restart the same process but now the probability to have a value $+1$ is $p_1$ for all the five spins. That is the second stroke, which in turn yields $p_2$ for the central spin when updated as shown in Figure (\ref{tt1}).
 
\begin{figure}[h]
\centering
\includegraphics[width=1\textwidth]{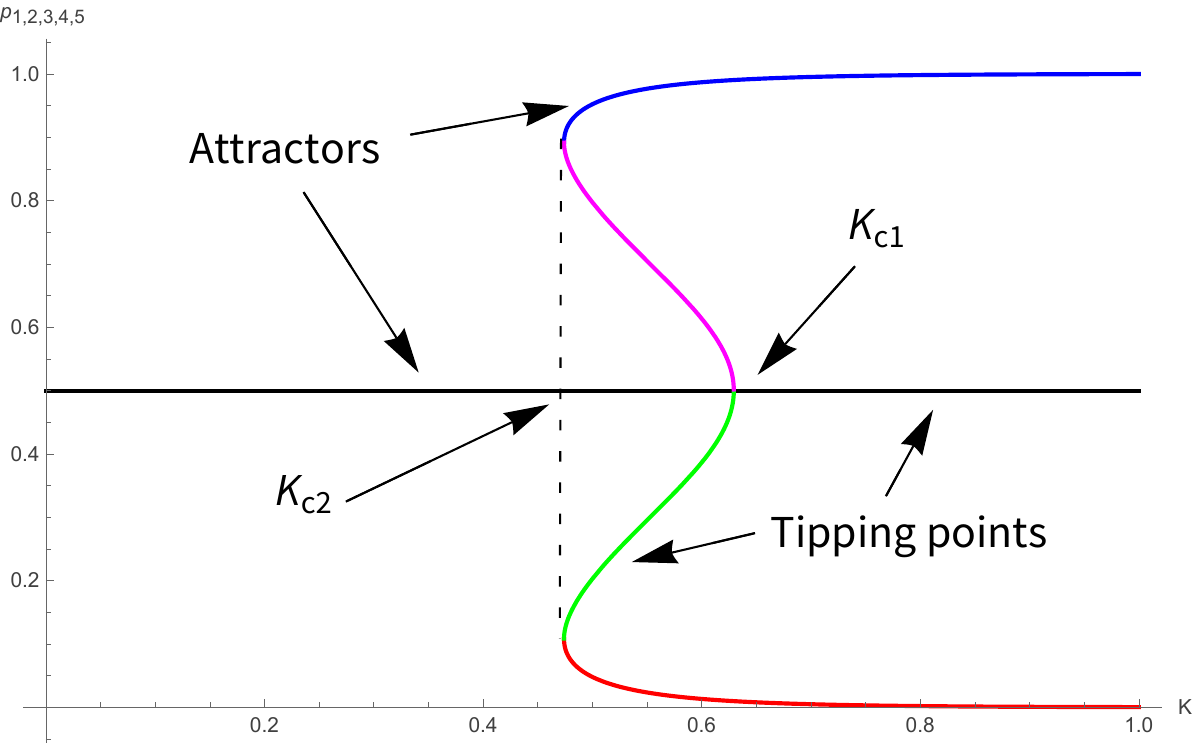}
\caption{Five fixed points $p_{1,2,3,4,5}$ from Eq. (\ref{pp4}) as a function of $K$. For $0 < K < K_{c1}$  the fixed point $1/2$ is attractor while it turns to a tipping for $K> K_{c1}$. For $K<K_{c2}$ the phase is always disordered. Upper blue and lower red parts are attractors while magenta and green parts are tipping points.}
\label{p2}
\end{figure}

Iterating the two stroke  pumping (TSP) technique $(n+1)$ times leads to a probability $p_{n+1}$ to have the central spin equal to $+1$ with,
\begin{eqnarray}
\label{p4} 
p_{n+1}&=&p_n \Bigl \{(1 - a^2) p_n^4 + (1 - a ) 4 p_n^3 (1 - p_n)  \Bigl \} \nonumber \\ 
&+& (1 - p_n)  \Bigl \{p_n^4 +  4 p_n^3 (1 - p_n)  + 6 p_n^2 (1 - p_n)^2 \Bigl \} \nonumber \\
&+ &   (1 - p_n) \Bigl \{ 4 a p_n (1 - p_n)^3 +  a^2 (1 - p_n)^4   \Bigl \} ,
\end{eqnarray}
where $p_n$ is the probability to have the central spin equal to $+1$ prior to the last update, $a\equiv \exp(-4 K)$ and $K\equiv J/T$.
Eq. (\ref{p4}) can be reduced to,
\begin{eqnarray}
\label{pp4} 
p_{n+1}&=&(1-a^2)  p_n^5 + (5 - 4a)  p_n^4(1 - p_n)  \nonumber \\ 
&+&4p_n^3(1-p_n)^2+ 6 p_n^2 (1 - p_n)^3  \nonumber \\
&+ &  4 a p_n (1 - p_n)^4+ a^2 (1 - p_n)^5 .
\end{eqnarray}
whose fixed points are shown in Figure (\ref{p2}). The associated transition is first order with $K_{c1}=0.629$ and $K_{c2}=0.474$, which is close to the Onsager value $K_O=0.441$.

\begin{figure}[h]
\vspace{-3cm}
\centering
\includegraphics[width=0.48\textwidth]{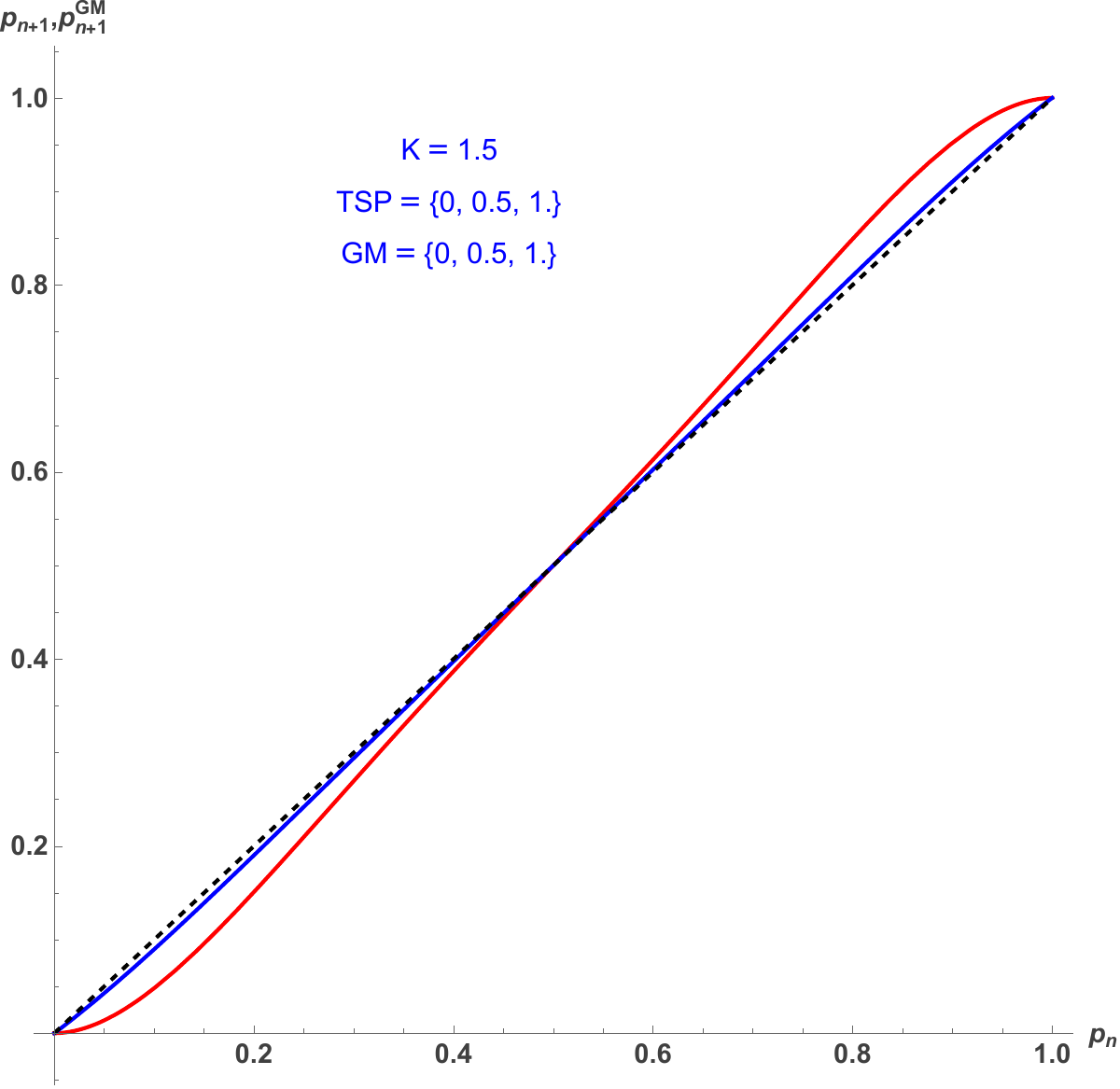}
\includegraphics[width=0.48\textwidth]{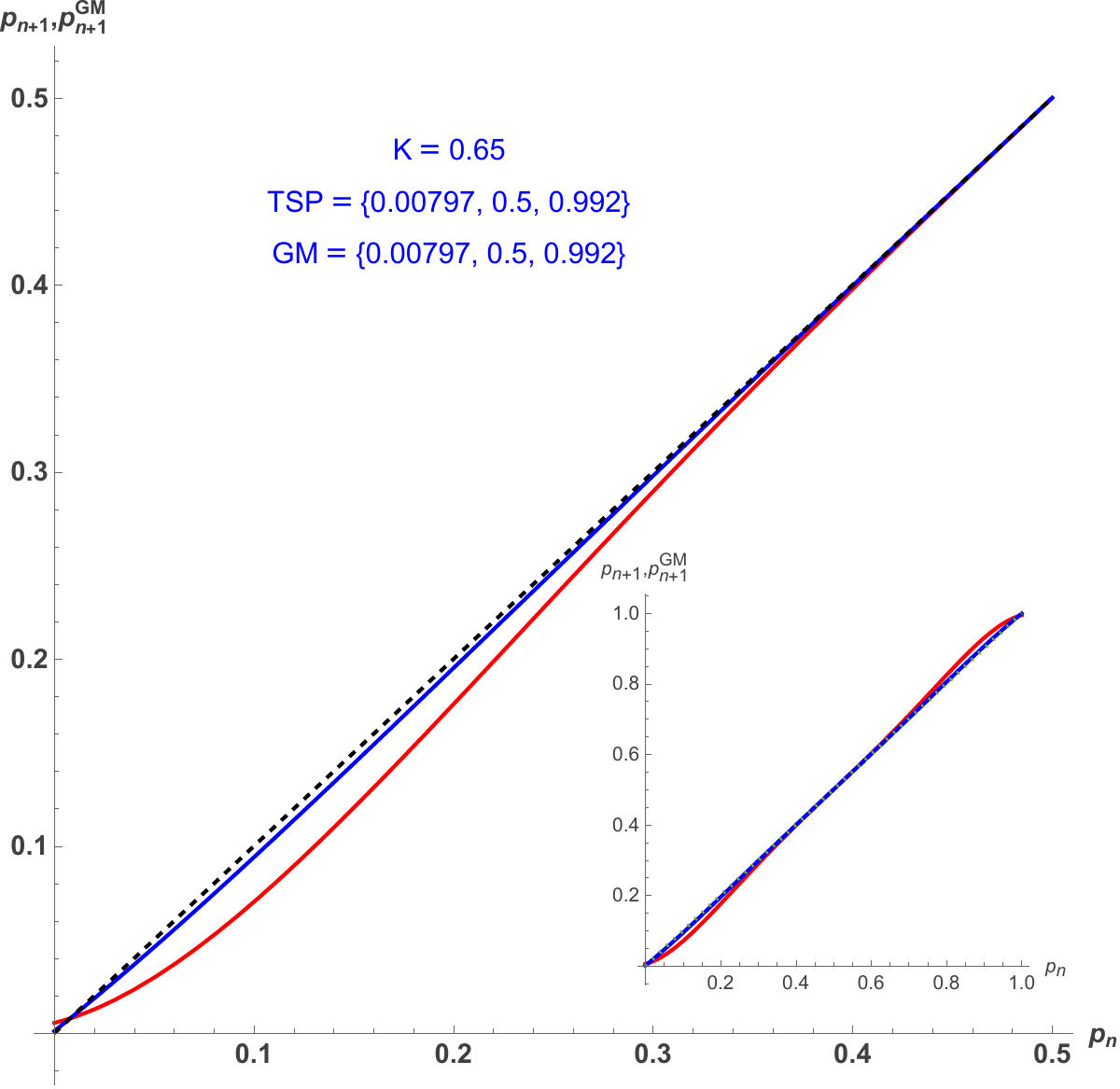}\\
\includegraphics[width=0.48\textwidth]{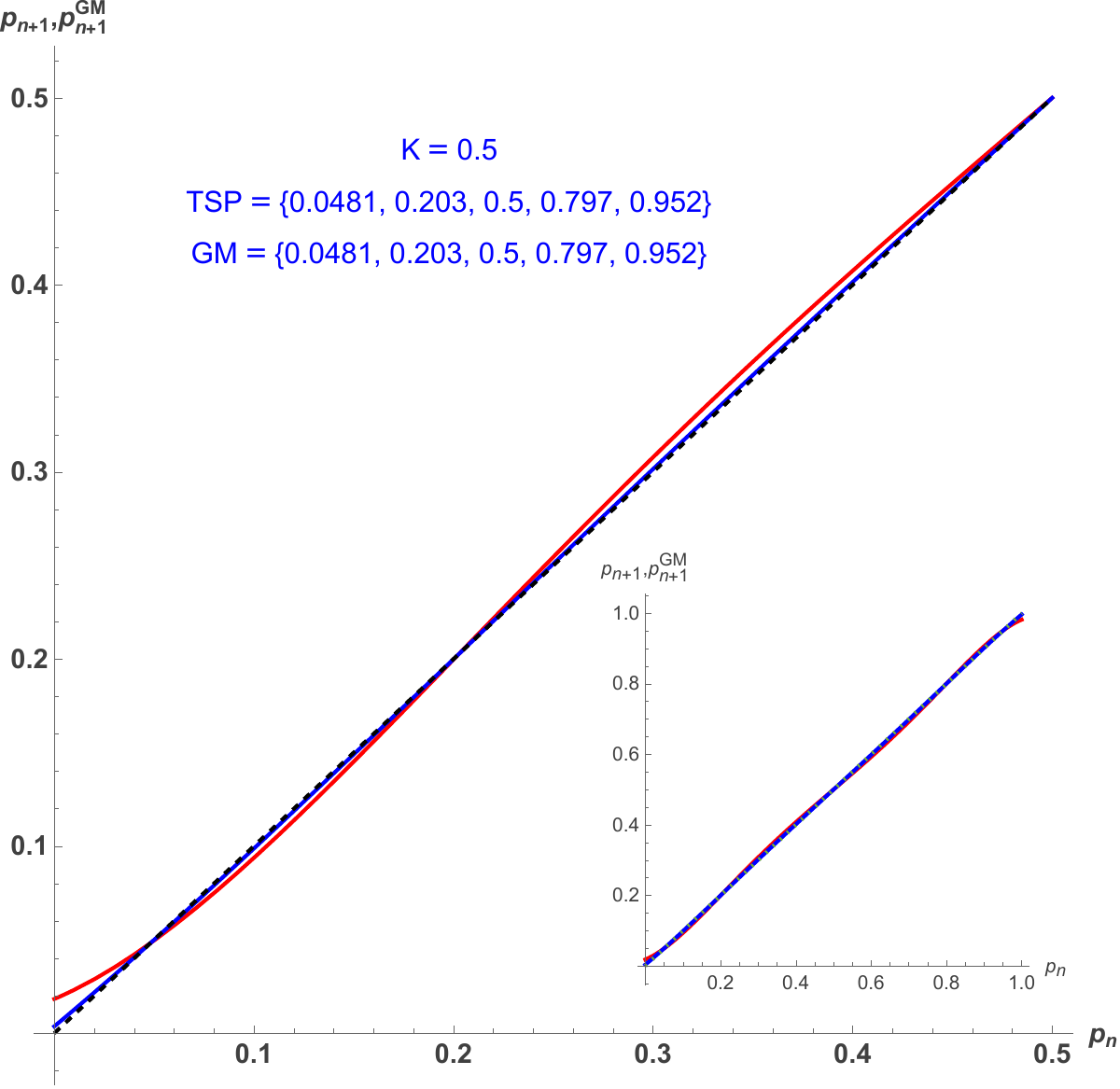}
\includegraphics[width=0.48\textwidth]{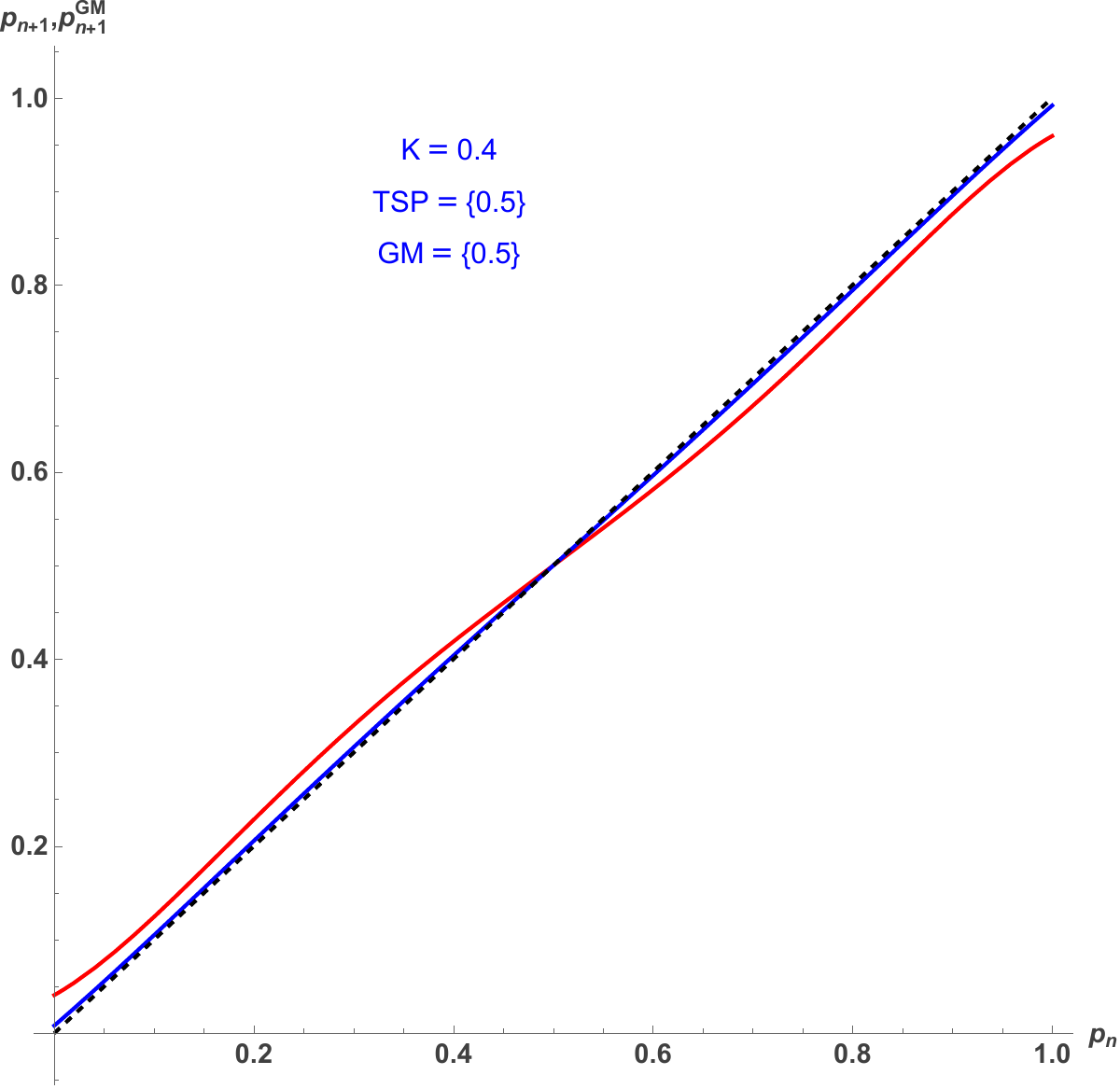}
\caption{Update probabilities $p_{n+1}$ and $p^{GM}_{n+1}$ given respectively by Eqs. (\ref{pp4}) in red and Eq. (\ref{m5}) in blue, as a function of $p_n$ for $K=1.5$ (upper left), $K=0.65$ (upper right), $K=0.5$ (lower right), $K=0.4$ (lower right). TSP and GM are the associated fixed points, which are found to be identical.}
\label{t1-4}
\end{figure}

\begin{figure}[h]
\vspace{-3.5cm}
\centering
\includegraphics[width=0.48\textwidth]{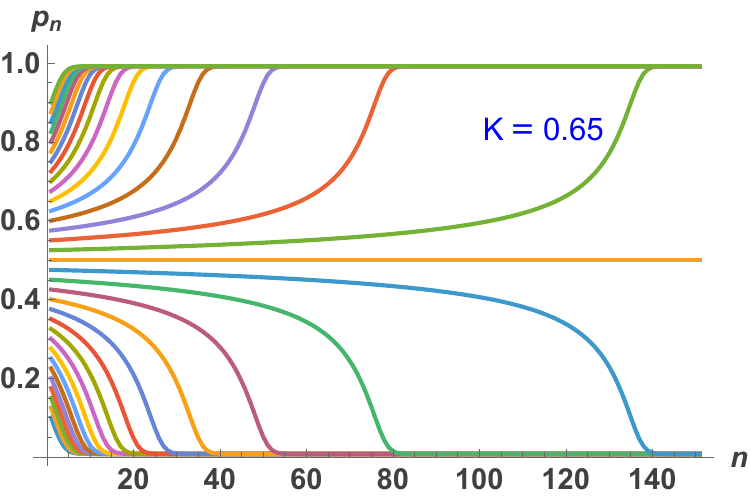}
\includegraphics[width=0.48\textwidth]{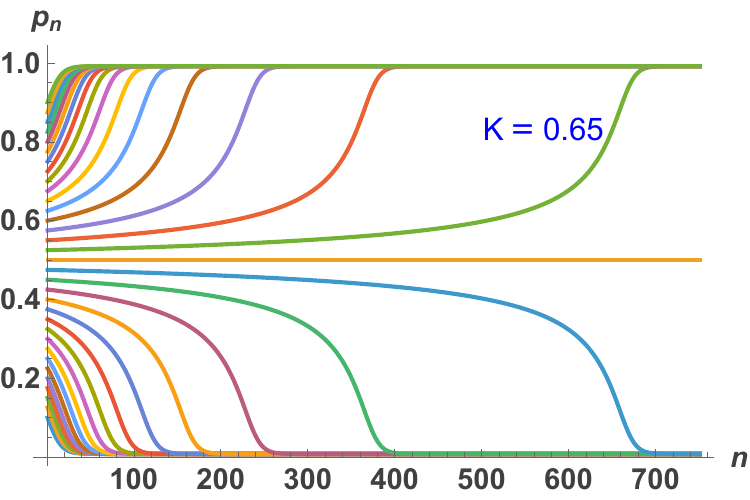}\\
\includegraphics[width=0.48\textwidth]{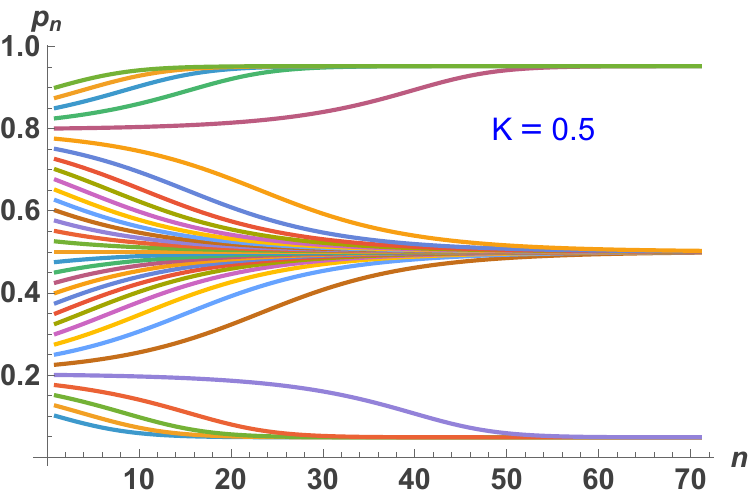}
\includegraphics[width=0.48\textwidth]{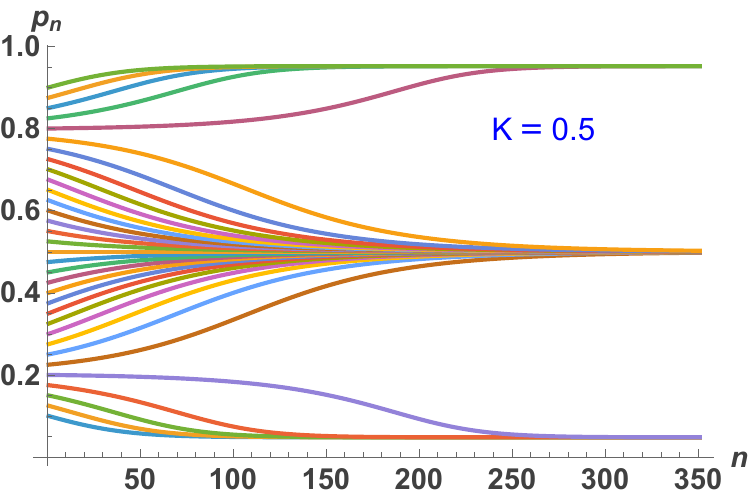}\\
\includegraphics[width=0.48\textwidth]{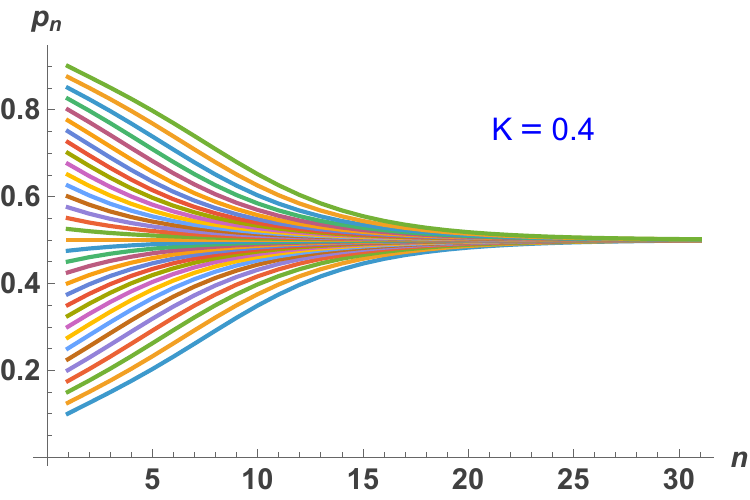}
\includegraphics[width=0.48\textwidth]{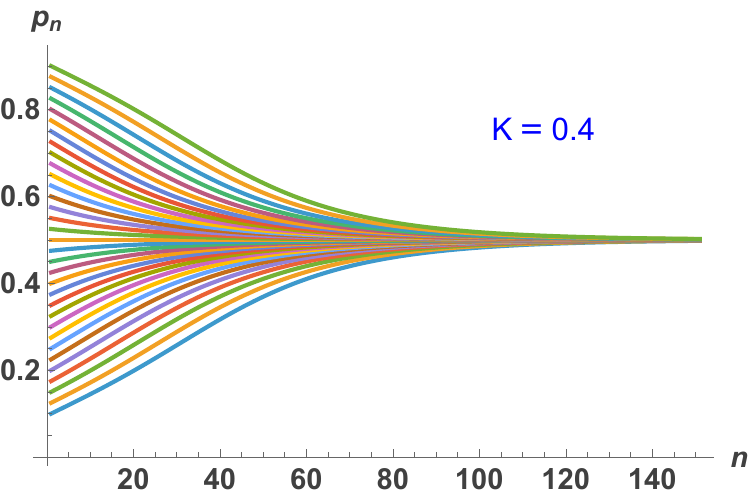}
\caption{Iterated updates from a series of initial values $p_0$ covering the full spectrum of values from 0 to 1, using Eqs. (\ref{pp4}) (left side) and (\ref{m5}) (right side). Three values of $K$ are shown with $K=0.65$ illustrating the case of full ordering (low temperatures), $K=0.5$ illustrating the case of the first oder region with either ordering or disorder as a function of the initial value $p_0$, $K=0.4$ illustrating the case of full disorder (high tempratures).}
\label{t5-10}
\end{figure}

With this respect, update equation is \cite{GalamMartins},
\begin{eqnarray}
\label{m5}
p^{GM}_{n+1}&=& \frac{1}{5} [( 5-a^2) p_n^5  +(21-4a) p_n^4 (1-p_n) \nonumber\\
&+&28 p_n^3 (1-p_n)^2+22 p_n^2 (1-p_n)^3  \nonumber\\
&+&  4 (1+a) p_n(1-p_n)^4 + a^2 (1-p_n)^5] ,
\end{eqnarray}
which is different from Eq. (\ref{pp4}). However, both update equations yield identical fixed points and first order transition as a function of $K$. However, while updates includes 5 spins, only one of them is updated, making slower the dynamics to reach the various attractors as seen in Figures (\ref{t1-4}) and (\ref{t5-10}). Figure (\ref{t5-10}) shows that about five more updates are required to reach the relevant attractor in comparison to TSP, which considers one spin at a time. 

More precisely, Figure (\ref{t5-10}) presents iterated updates from a series of initial values $p_0$ covering the full spectrum of values from 0 to 1, using Eqs. (\ref{pp4}) (left side) and (\ref{m5}) (right side). Three values of $K$ are shown with $K=0.65$ illustrating the case of full ordering (low temperatures), $K=0.5$ illustrating the case of the first oder region with either ordering or disorder as a function of the initial value $p_0$, $K=0.4$ illustrating the case of full disorder (high temperatures). 

\section{The 1-dimensional case}

\begin{figure}
\centering
\vspace*{-6.5cm}
\hspace*{-1.5cm}
\includegraphics[width=1.25\textwidth]{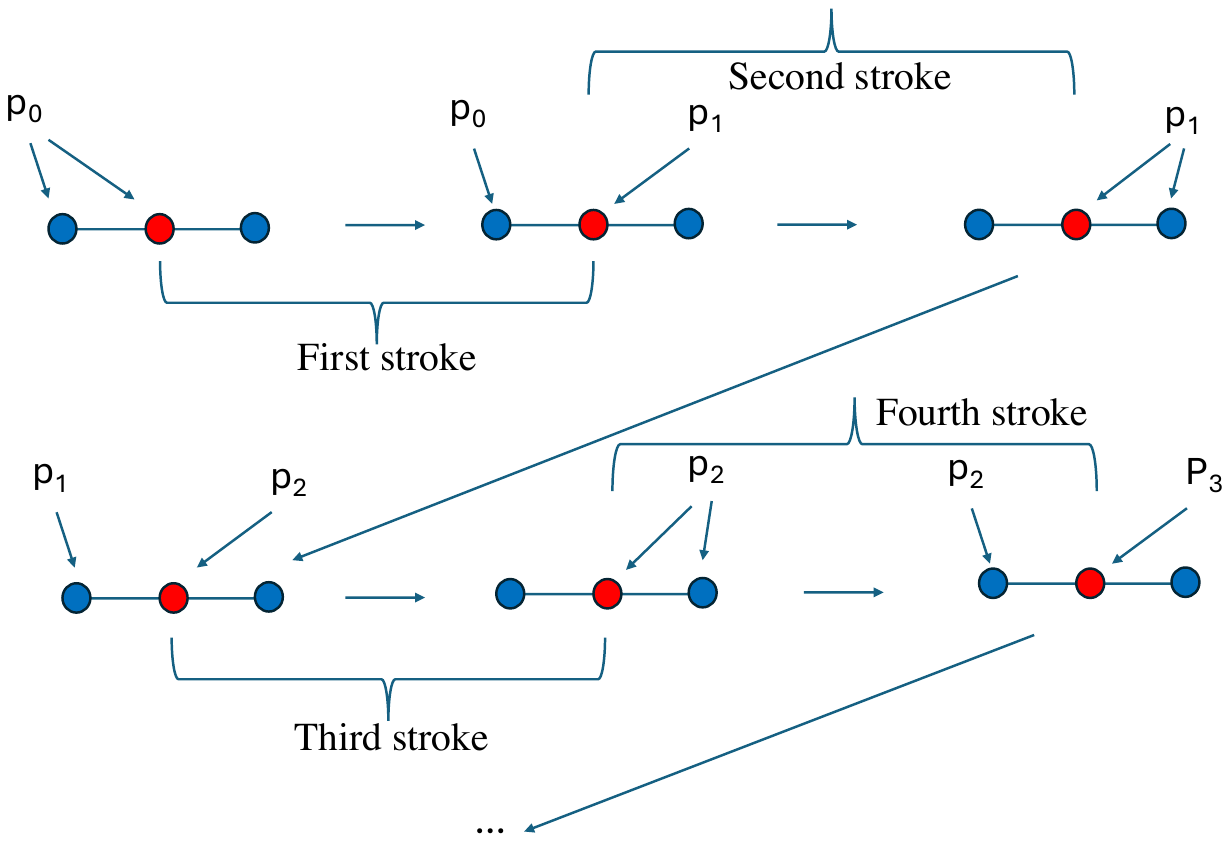}
\caption{Up left: a cluster of one spin and its 2 nearest-neighbors with probability $p_0$ to have $+1$ for each of them.  Up center: The central spin has been updated with a probability $p_1$ to be $+1$. That defines the first stroke of the pumping technique. Up right: the probability to $+1$ is now $p_1$ for all 3 spins defining the second stroke. Down left: the central spin is updated with probability $p_2$ to be $+1$. Down center: The probability to be $+1$ is $p_2$ for all 3 spins. Down right: the central spin is updated with probability $p_3$ to $+1$. and so forth till an attractor is reached.}
\label{tt2}
\end{figure}

I now apply TSP to the 1-dimensional Ising model where each spin has two nearest-neighbors. Figure (\ref{tt2}) illustrate the associated dynamics. The resulting update function is, 

\begin{equation}
p_{n+1}=p_n (1 - a) p_n^2 +  (1 - p_n) (p_n^2 + 2 p_n (1 - p_n) + a (1 - p_n)^2)   ,
\label{p1}
\end{equation}
which reduce to,
\begin{equation}
p_{n+1}= (1 - a) p_n^3+  (1 - p_n) p_n^2 + 2 p_n (1 - p_n)^2 + a (1 - p_n)^3   .
\label{pp1}
\end{equation}

\begin{figure}[h]
\vspace{-2cm}
\centering
\includegraphics[width=0.48\textwidth]{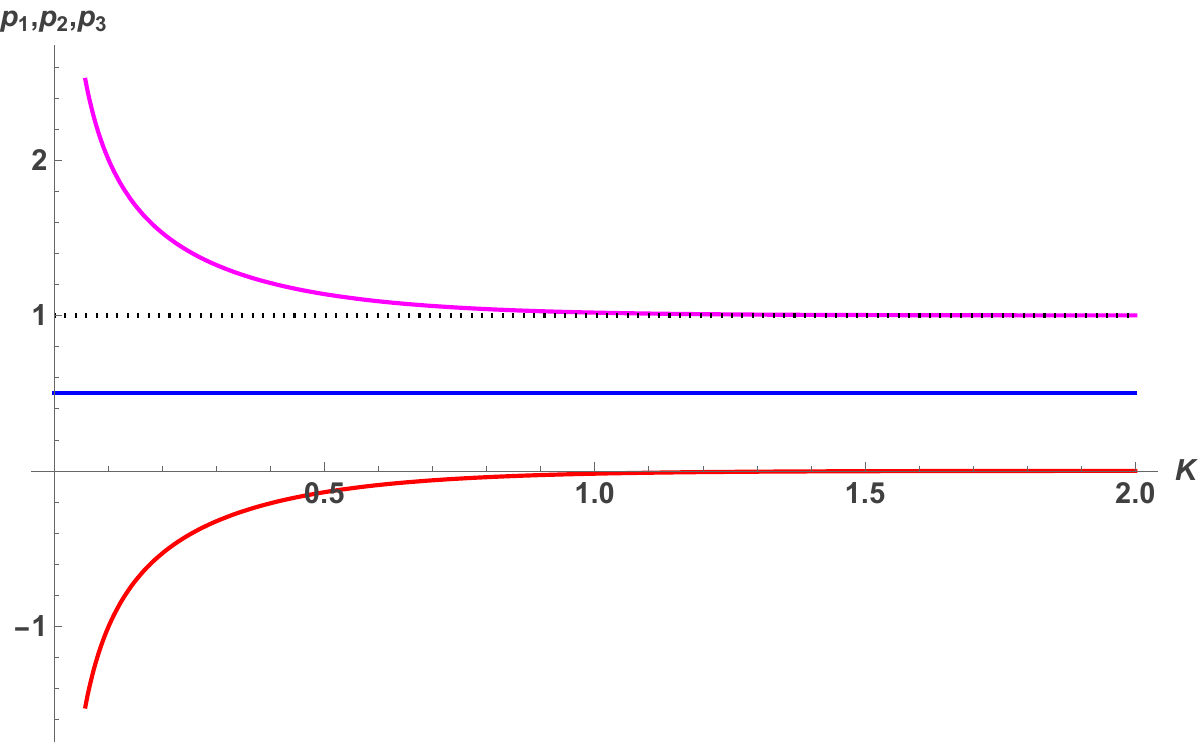}
\includegraphics[width=0.48\textwidth]{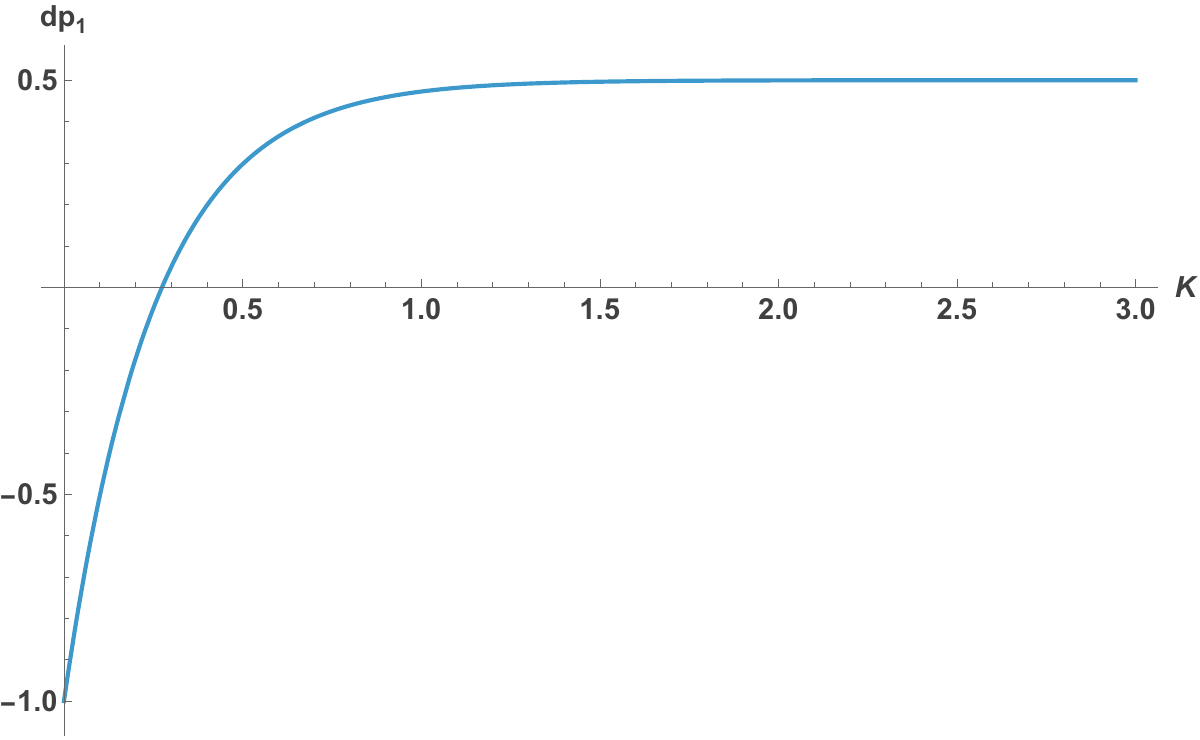}
\caption{Fixed points of Eq. (\ref{pp1}) as a function of $K$. Only $p_1=1/2$ is valid with the asymptotic limits $0,1$ of $p_{2,3}$, i.e., at $T=0$. }
\label{u12}
\end{figure}

Solving Eq. (\ref{pp1}) fixed point equation yields 3 values $p_1, p_2, p_3$ with $p_1=1/2$ and,
\begin{equation}
p_{2,3}=\frac{-1+a^{-1}\pm\sqrt{-3+2a^{-1}+a^{-2}}}{2(-1+a^{-1})} ,
\label{p23}
\end{equation}
which are shown in left part of Figure (\ref{u12}) as a function of $K$. Only $p_1=1/2$ is valid with the asymptotic limits $0,1$ of $p_{2,3}$. 

The derivative of Eq. (\ref{pp1}) taken at $p=1/2$ is,
\begin{equation}
dp_{1}=-1+\frac{3}{2}(1-a) ,
\label{dpp}
\end{equation}
which satisfies $-1 \leq dp_{1} \leq 1/2$ as seen in the right part of Figure (\ref{u12}) indicating $p_1=1/2$ is always attractor for any value of $K$ including its infinite limit, i.e., $T=0$.

\begin{figure}[h]
\centering
\includegraphics[width=0.48\textwidth]{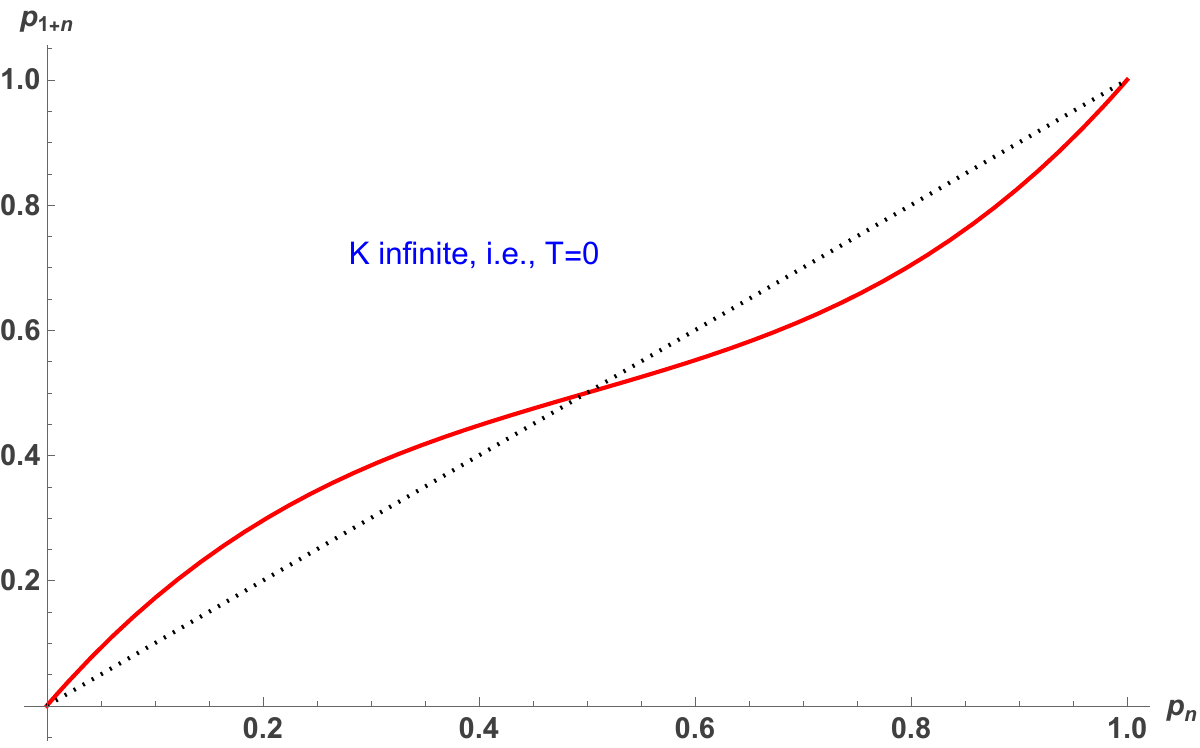}
\includegraphics[width=0.48\textwidth]{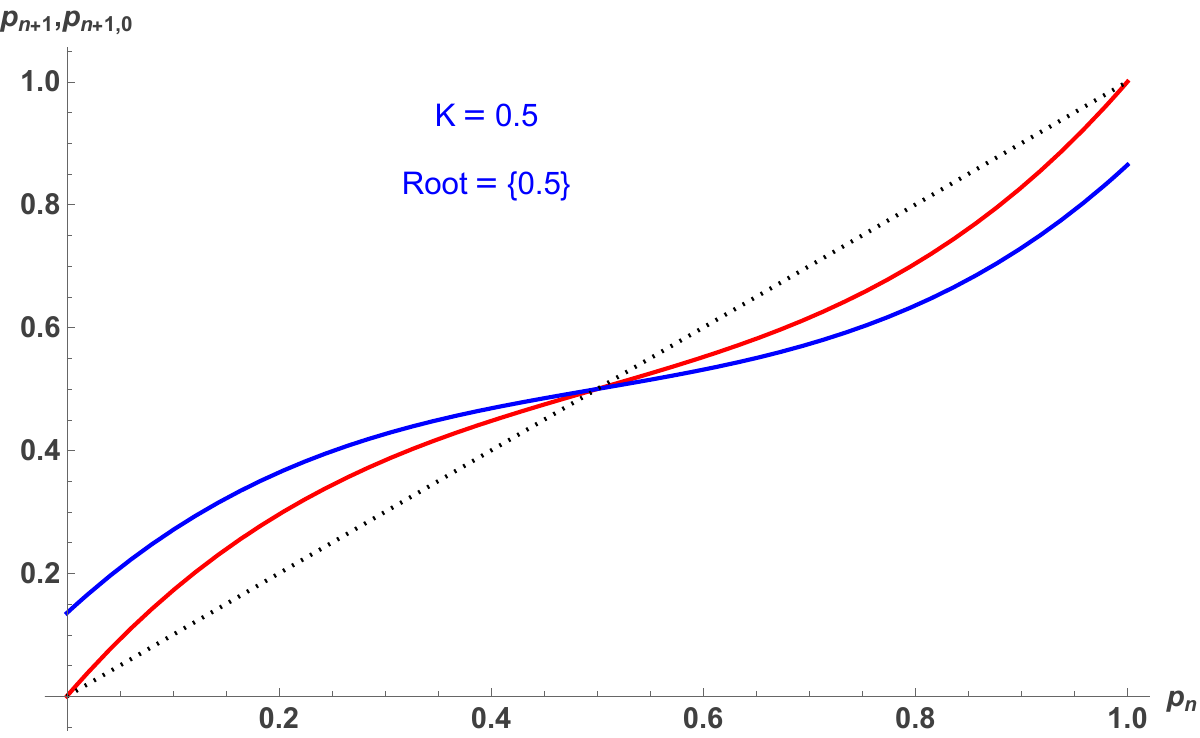}
\caption{Eq. (\ref{pp10}) is shown in the part. Right part shows Eq. (\ref{pp10}) and Eq. (\ref{pp1}) for $K=0.50$.}
\label{u34}
\end{figure}

Indeed, at $T=0 \Leftrightarrow a=0$ Eq. (\ref{pp1}) becomes,
\begin{equation}
p_{n+1}= 2p_n^3 -3  p_n^2 + 2 p_n   ,
\label{pp10}
\end{equation}
which yields three fixed points $(0,1/2,1)$ where $p=1/2$ is still attractor and $(0,1)$ two tipping points. Eq. (\ref{pp10}) is shown in  the left part of Figure (\ref{u34}). Right part of the Figure shows Eq. (\ref{pp10}) and Eq. (\ref{pp1}) for $K=0.50$.

Therefore, $p=1/2$ is always attractor even at $T=0$. However at $T=0$, while the values $0,1$ are unstable fixed points, as soon as $T\neq 0$ they stop to be fixed points as expected from the fact that the 1-dimensional Ising model has no ordered phase at $T\neq 0$.

At this stage, the TSP technics yields a dynamics driven by two repulsive fixed points at $p=0$ and $p=1$ separated by an attractor located at $p=1/2$. However, the one dimensional Ising model has a repulsive fixed point at $p=1/2$ with two attractors at respectively $p=0$ and $p=1$.

While this apparent contrast could, at first sight, be interpreted as a failure of the TSP technics, it acquires a different meaning once the notion of dynamical stability is evoked. In particular, discriminating asymptotic properties and operationally dynamic results sheds a new light on the TSP outcome.

With this respect, it is of importance to notice that at $T=0$ and $d=1$ the single-spin-flip dynamics produces diffusive domain-wall motion without energy change, leading to relaxation times diverging as $t_{\mathrm{ord}} \sim L^{2}$ \cite{krapivsky}. 

As a consequence, although the ground state is asymptotically reached, the evolution of the magnetization becomes arbitrarily slow and no full ordering is achieved on finite, physically or numerically accessible time scales.

This behavior is directly consistent with what is observed in Monte Carlo simulations, which necessarily probe finite times and finite system sizes. For instance system sizes $L=10^4$ and  $L=10^5$ correspond to ordering times $t \sim 10^8$ and $t \sim 10^{10}$ respectively, which are out of reach for standard Monte Carlo studies. 

From this perspective, the TSP dynamics could be seen as providing an effective description of the dynamical states that are actually reached in simulations, whereas the asymptotic fixed-point structure of the 1d Ising model becomes relevant on diverging time scales. The origin of this property can be connected to the use of the single spin Metropolis update in the TSP technics.

Therefore, TSP outcome is not to contrast the asymptotic fixed-point structure where $p=1/2$ is a repeller, but rather to contribute an additional insight on the actual dynamics observed on finite, physically and numerically accessible time scales.

With this regard, it is worth stressing that RG transformations eliminate dynamically generated absorbing states by the related coarse-graining and thus do not yield above situation. As a result, the RG description reproduces the correct equilibrium fixed points but fails to account for the dynamics that makes the ordered state practically inaccessible from a generic initial condition. 

Likewise, Bethe approximation yields the equilibrium ordered phases but fails to account for the slowing done dynamical that arises when the system relaxes under local T = 0 dynamics from a mixed initial condition within finite times. 

\section{The 3-dimensional case}

\begin{figure}[h]
\centering
\includegraphics[width=0.8\textwidth]{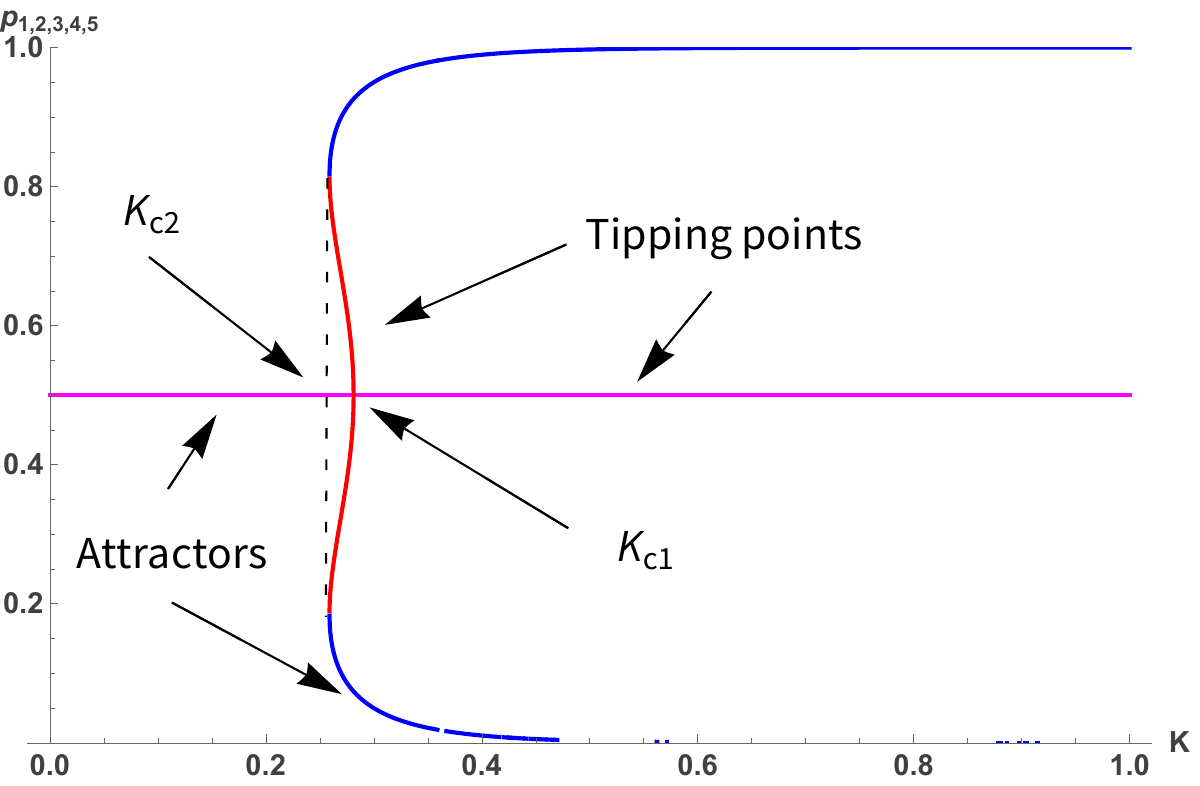}
\caption{Five fixed points $p_{1,2,3,4,5}$ from Eq. (\ref{p6}) as a function of $K$. For $0 < K < K_{c1}$  the fixed point $1/2$ is attractor while it turns to a tipping for $K> K_{c1}$. For $K<K_{c2}$ the phase is always disordered. Upper and lower blue parts are attractors while red parts are tipping points.}
\label{u5}
\end{figure}

Applying TSP to the 3-dimensional Ising model yields the update equation, 
\begin{eqnarray}
\label{p6}
p_{n+1}&=& (1 - a^3 p_n) p_n^6 + 6(1 - a^2 p_n)  p_n^5 (1 - p_n) \nonumber\\
&+& 15 (1 - a p_n) p_n^4 (1 - p_n)^2  + 20 p_n^3 (1 - p_n)^4 \nonumber\\
&+ & 15 a  p_n^2 (1 - p_n)^5  + 6 a^2  p_n (1 - p_n)^6  + a^3 (1 - p_n)^7 ,
\end{eqnarray}
whose fixed points are exhibited in Figure(\ref{u5}).

\begin{figure}[h]
\vspace{-2cm}
\centering
\includegraphics[width=0.48\textwidth]{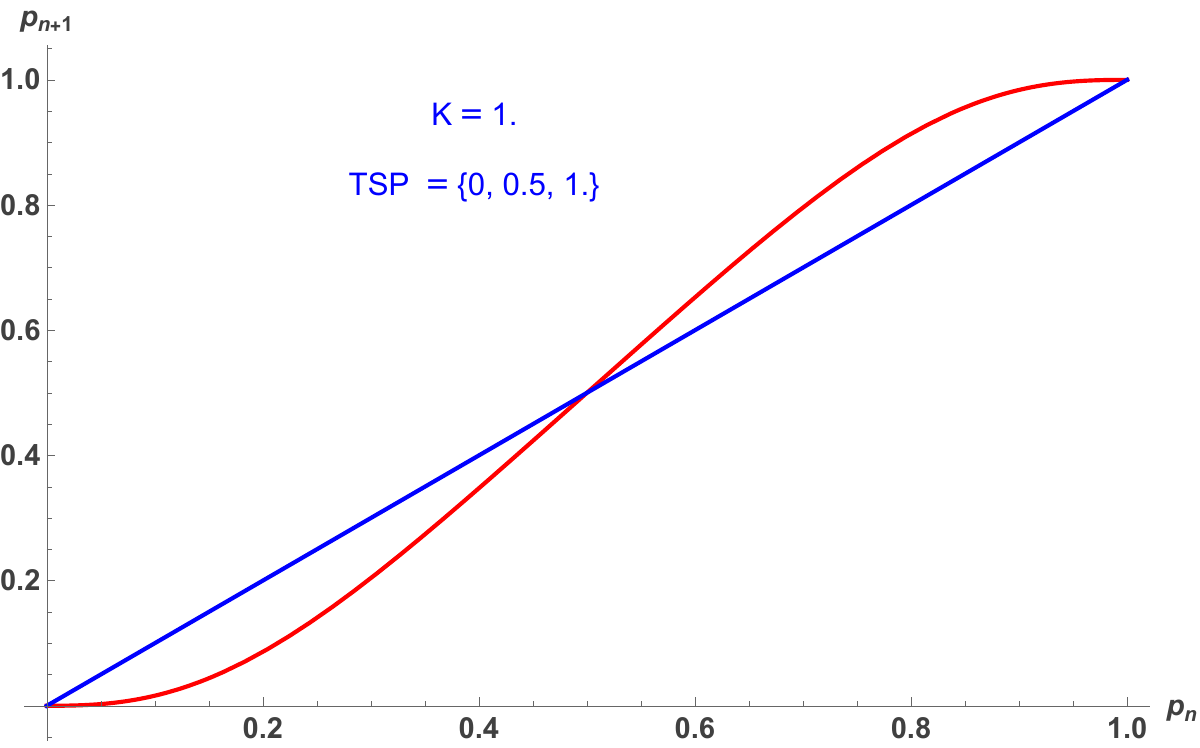}
\includegraphics[width=0.48\textwidth]{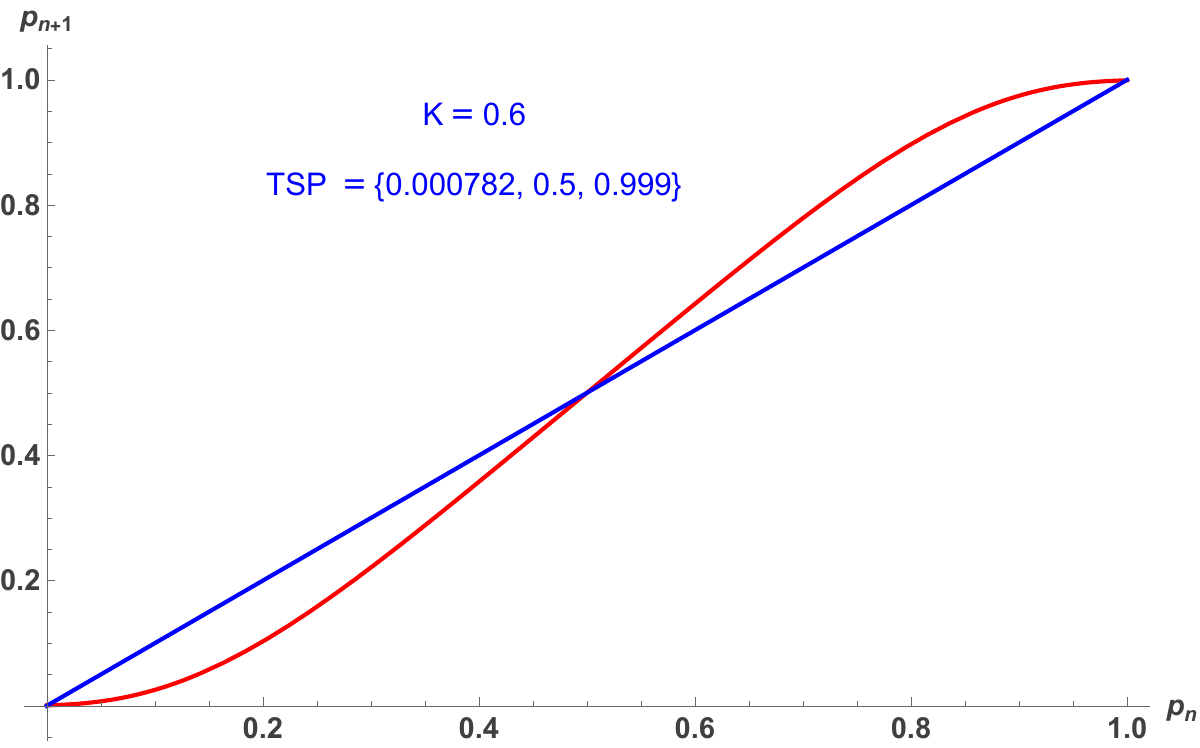}\\
\includegraphics[width=0.48\textwidth]{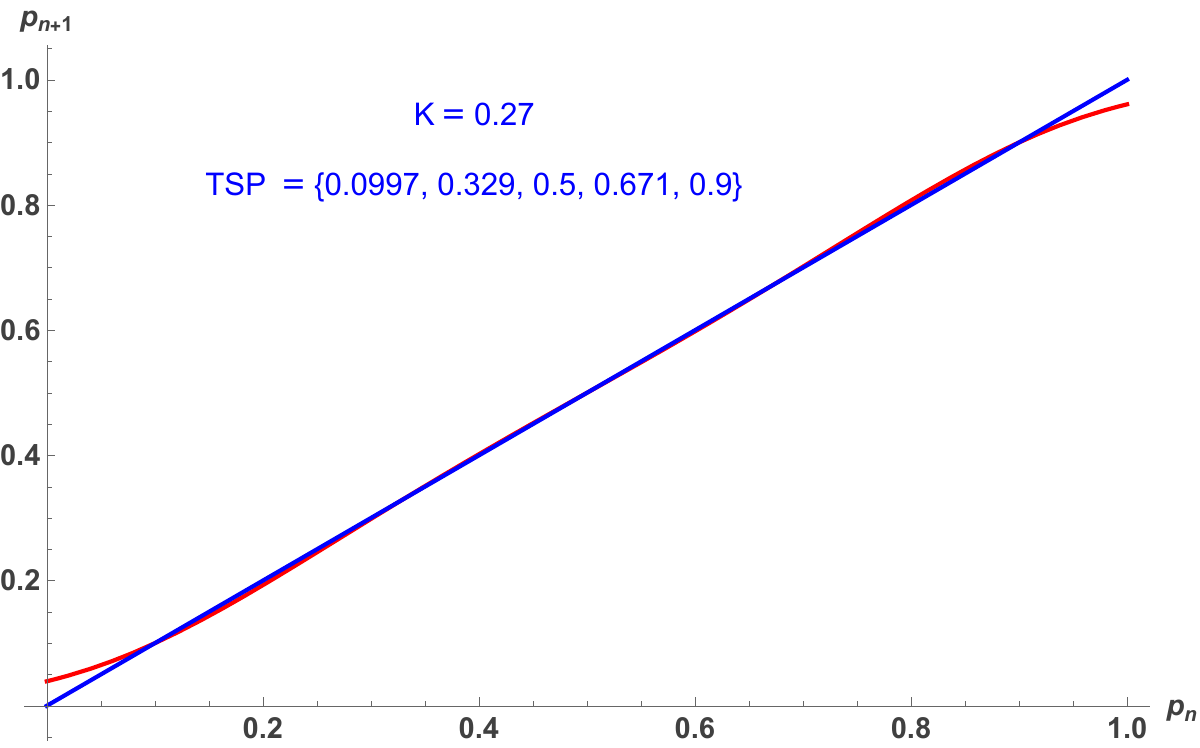}
\includegraphics[width=0.48\textwidth]{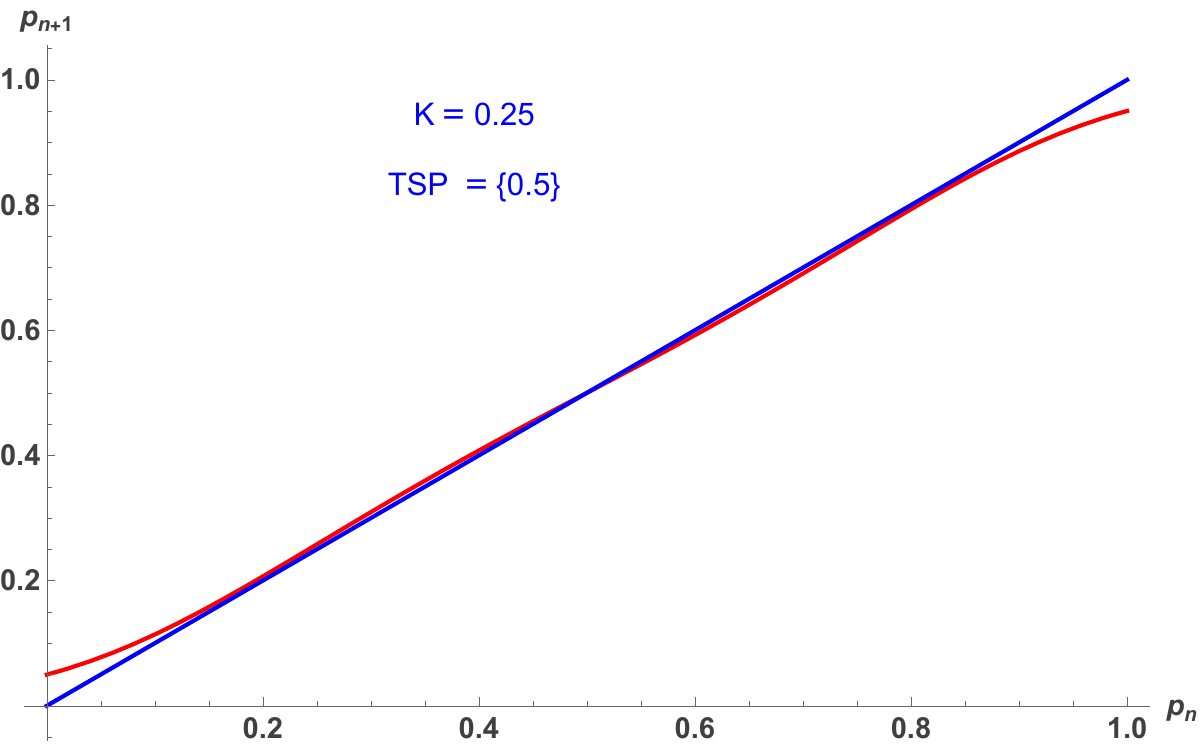}
\caption{Update probabilities $p_{n+1}$ given  by Eqs. (\ref{p6}) in red as a function of $p_n$ for $K=1$ (upper left), $K=0.6$ (upper right), $K=0.27$ (lower right), $K=0.25$ (lower right). Respective fixed points are indicated.}
\label{u6-8}
\end{figure}

While the phase diagram is similar to the 2-dimensional case, the extend of the first order transition is reduced with now $K_{c1}=0.280$ and $K_{c2}=0.255$. The same applies to Eq. (\ref{p6}) as seen from Figure (\ref{u6-8}) with respect to Figure (\ref{t1-4}).

\section{The 4-dimensional case}

\begin{figure}[h]
\centering
\includegraphics[width=0.8\textwidth]{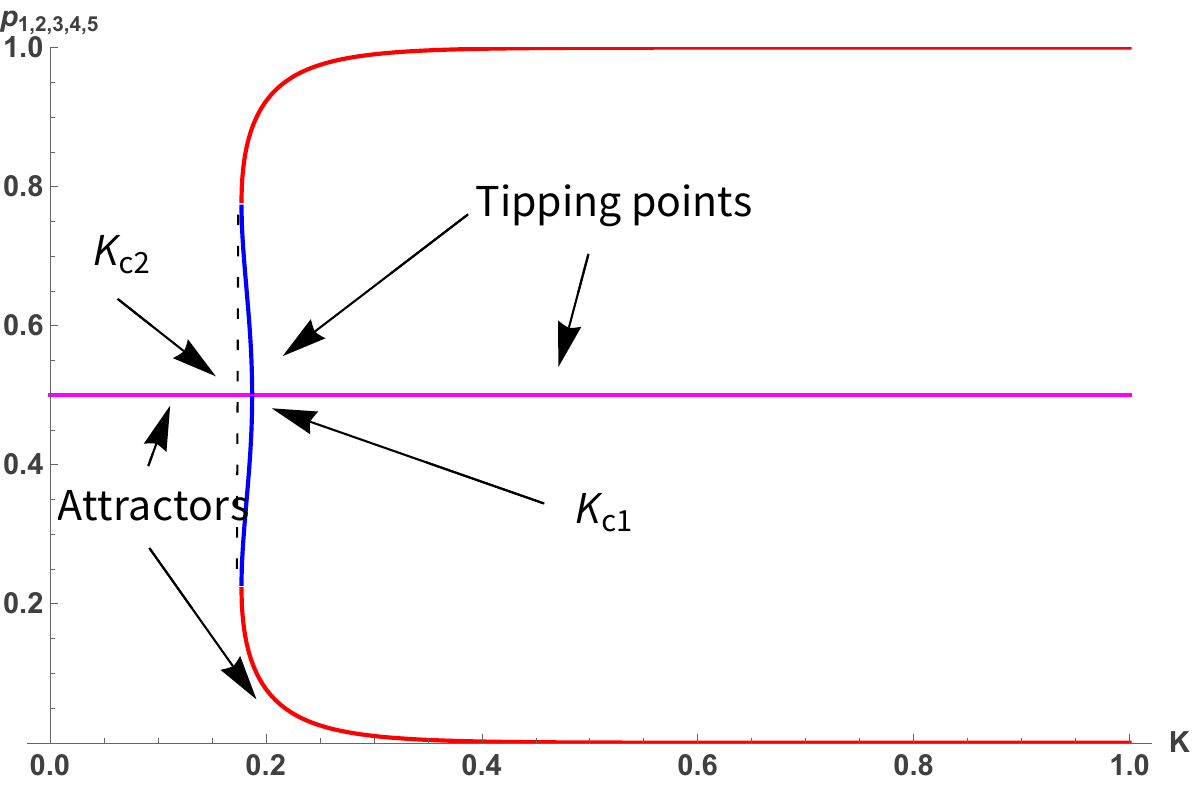}
\caption{Five fixed points $p_{1,2,3,4,5}$ from Eq. (\ref{p8}) as a function of $K$. For $0 < K < K_{c1}$  the fixed point $1/2$ is attractor while it turns to a tipping for $K> K_{c1}$. For $K<K_{c2}$ the phase is always disordered. Upper and lower blue parts are attractors while red parts are tipping points.}
\label{v1}
\end{figure}

At $d=4$ the update equation becomes,
\begin{eqnarray}
\label{p8}
p_{n+1}&=& (1 - a^4 p_n) p_n^8 + 8 (1 - a^3 p_n) p_n^7 (1 - p_n) + 28 (1 - a^2 p_n)  p_n^6 (1 -  p_n)^2 \nonumber\\
&+& 56(1 - a p_n) p_n^5 (1 - p_n)^3 + 70 p_n^4 (1 - p_n)^5 + 56 a p_n^3 (1 - p_n)^6  \nonumber\\
&+ & 28 a^2 p_n^2 (1 - p_n)^7 + 8 a^3  p_n (1 - p_n)^8 + a^4 (1 - p_n)^9 ,
\end{eqnarray}
with its fixed points shown in Figure (\ref{v1}). 

\begin{figure}[h]
\vspace{-2cm}
\centering
\includegraphics[width=0.48\textwidth]{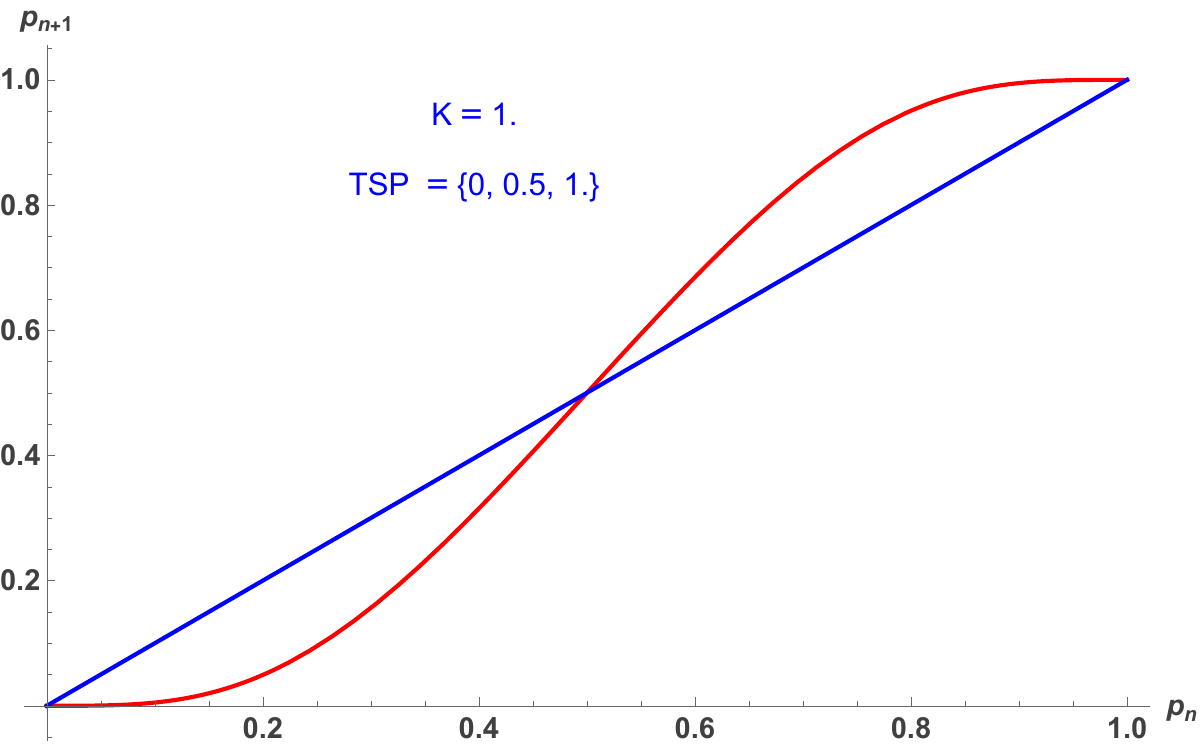}
\includegraphics[width=0.48\textwidth]{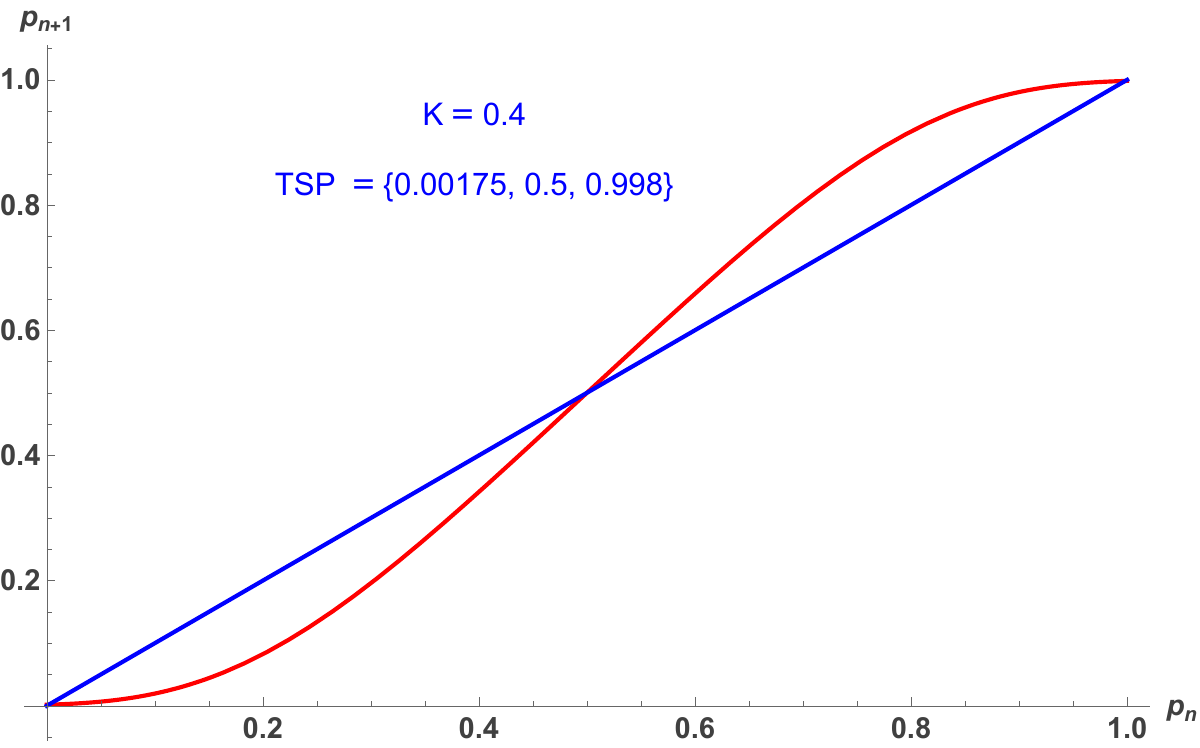}\\
\includegraphics[width=0.48\textwidth]{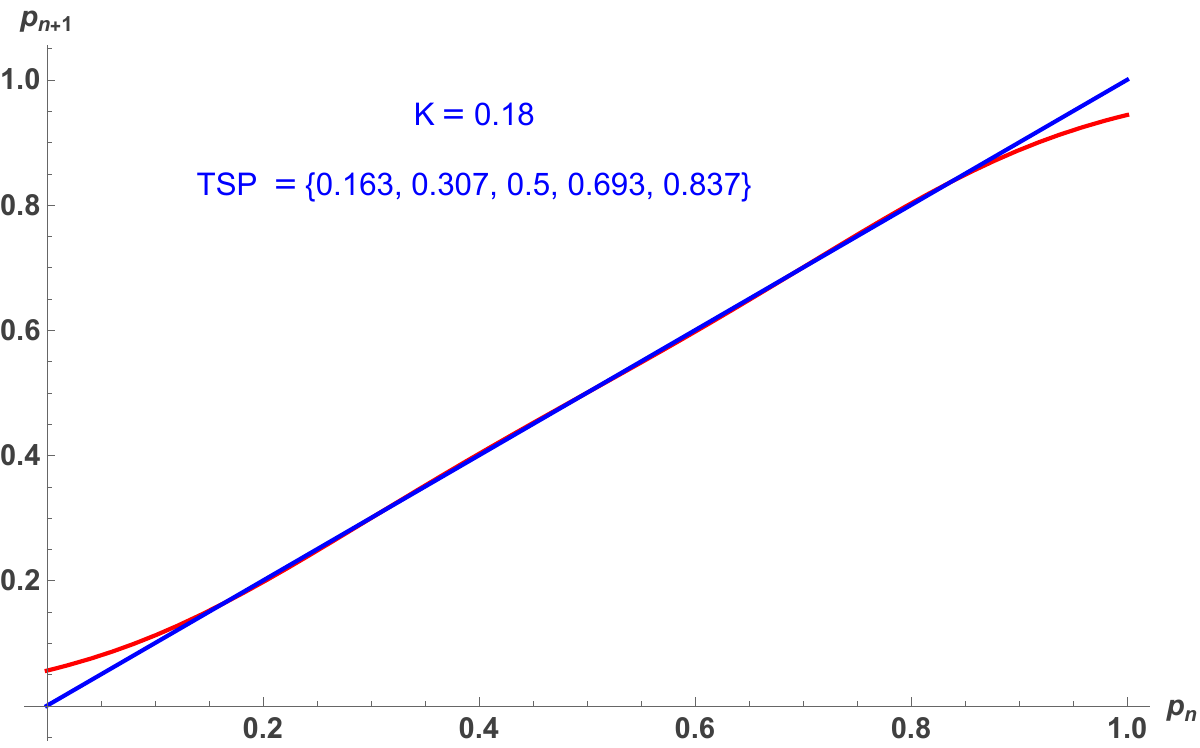}
\includegraphics[width=0.48\textwidth]{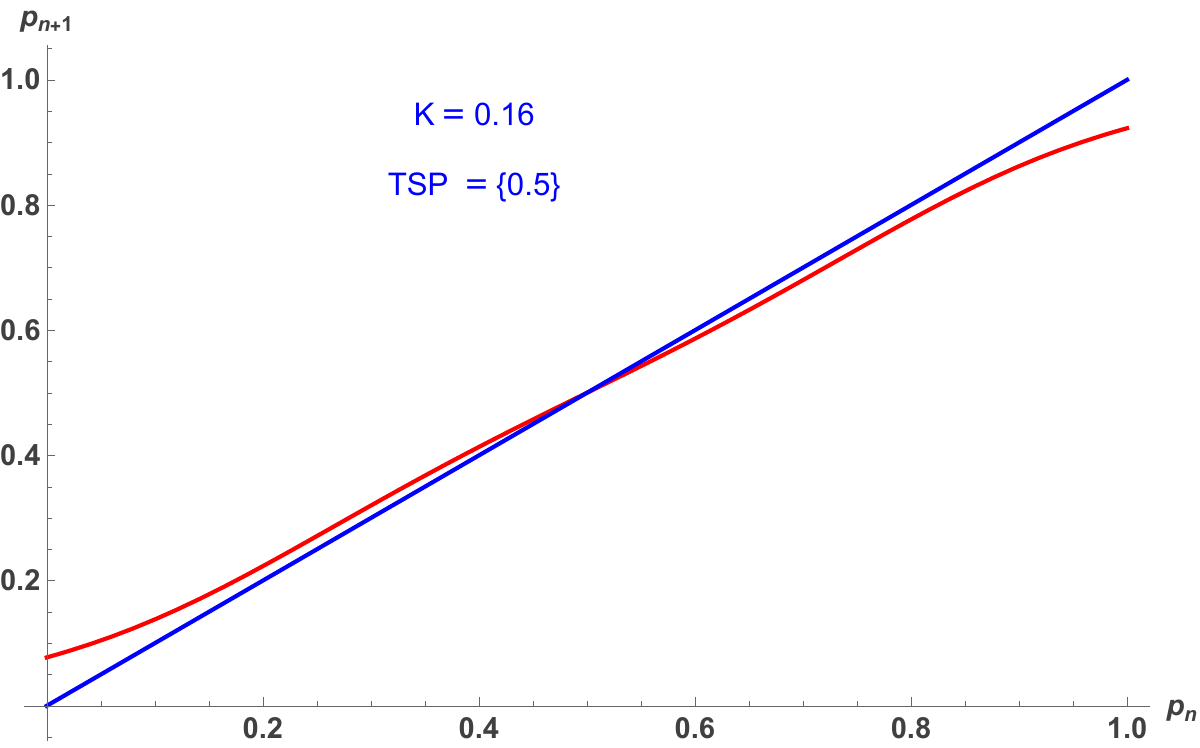}
\caption{Update probabilities $p_{n+1}$ given  by Eqs. (\ref{p8}) in red as a function of $p_n$ for $K=1$ (upper left), $K=0.4$ (upper right), $K=0.18$ (lower right), $K=0.16$ (lower right). Respective fixed points are indicated.}
\label{v2-5}
\end{figure}

As for moving from $d=2$ up to $d=3$, moving up to $d=4$ reduces against the first order region to a very narrow strip with $K_{c1}=0.185$ and $K_{c2}=0.175$. Figure (\ref{v2-5}) illustrate the shrinking.

\section{Conclusion}

In this paper I have  introduced a novel technique to evaluate the critical temperatures of the nearest-neighbors Ising model at any dimension $d$. Denoted two stroke pumping (TSP) I have implemented it at $d=1, 2, 3, 4$. Associated values of $K_c$ are exhibited in Table (\ref{TSP}) along with estimates using MF, Bethe, GM and MC. All values are reported with two digits only since this precision is sufficient for making meaningful comparisons.

TMS values overestimates exact estimates with a constant excess of $+0.03$. The fact that the difference is constant indicates it does not depend on dimension. However, at the moment I have no explanation of its origin. 

TSP and Bethe hold a series of respective advantages. Bethe (i) correctly predicts a second order phase transition at $d=2$; (ii)  is asymptotically exact in the $d\rightarrow \infty$ limit, and (iii) yields more precise values of $T_c$ at $d > 2$. In contrast, STP (i) includes the dynamics to reach equilibrium; (ii) at $d=1$ shows that the full ordering at $T=0$, it is not reachable on finite, physically and numerically accessible time scales, in agreement with MC simulations, while requiring significantly fewer computational resources; (iii) and yields a much better value of $T_c$ at $d=2$, yet with a first order transition.

The STP framework is general, does not rely on lattice-specific geometric assumptions, and can be extended to a broad class of discrete spin models. It would interesting in a future work to apply this two stroke pumping technique to other spin and lattice models with complex coupling structures, diluted fields, quenched disorder, and other many-body systems.

\begin{table}[h]
\centering
\begin{tabular}{|c|c|c|c||c||c|c|c|c|}
\hline
$d$ & $z$ & MF & Bethe & TSP &GM & MC \\
\hline
1 & 2  & 0.50 & \(\infty\) & \(\infty\) &  \(\infty\) & \(\infty\) \\
2 & 4  & 0.25 (-0.19) & 0.35 (-0.09) & 0.47 (+0.03) & 0.44 & 0.44\\
3 & 6  & 0.17 (-0.05)& 0.20 (-0.02) & 0.25 (+0.03) & 0.22& 0.22\\
4 & 8  & 0.13 (-0.02)& 0.14 (-0.01) & 0.18 (+0.03)& 0.15 & 0.15 \\
\hline
\end{tabular}
\caption{Critical coupling $K_c = 1/T_c$ of the nearest-neighbor Ising model ($J=1$, $k_B=1$), rounded to 2 significant digits for dimensions $d=1, 2, 3, 4$ and a number of respective numbers of nearest-neighbors $z=2, 4, 6, 8$ using mean-field (MF); Bethe; Two stroke pumping (TSP); GM formula; Monte Carlo simulations (MC). Parentheses indicate the difference between each estimate with the MC one. GM yields MC values.}
\label{TSP}
\end{table}


\end{document}